\documentclass{icfi} 
 
\begin{document} 
\markboth{E. Elizalde} 
{COSMOLOGY: TECHNIQUES AND OBSERVATIONS} 
 
%
\catchline{}{}{}{}{} 
%
 
\title{COSMOLOGY: TECHNIQUES AND OBSERVATIONS\footnote{Lectures given at the ``II International 
Conference on Fundamental Interactions'', Pedra Azul, Brazil, June 
6-12, 2004.} } 
 
\author{\footnotesize Emilio Elizalde\footnote{ On leave from 
ICE, Consejo Superior de Investigaciones Cient\'{\i}ficas,  and 
IEEC, Edifici Nexus, Gran Capit\`{a} 2-4, 08034 Barcelona, Spain. 
E-mail: elizalde@math.mit.edu, elizalde@ieec.fcr.es.}} 
 
\address{Department of Mathematics, Massachusetts Institute of 
Technology \\ 77 Massachusetts Ave, Cambridge, MA 02139-4307, USA. 
} 
 
 
\maketitle 
 
 
\begin{abstract} 
These lectures were addressed to nonspecialists willing to learn 
some basic facts, approaches, tools and observational evidence 
which conform  modern cosmology.  The aim is also to try to 
complement the many excellent treatises that exists on the subject 
(an exhaustive treatment being in any case impossible for lack of 
time, in the lectures, and of space here), instead of trying to 
cover everything in a telegraphic way. We start by 
 recalling in the introduction a couple of philosophical 
questions that have always upset inquiring minds. We then present 
some original mathematical approaches to investigate a number of 
basic questions, as the comparison of two point distributions 
(each point corresponding to a galaxy or galaxy cluster), the use 
of non-standard statistics in the analysis of possible 
non-Gaussianities, and the use of zeta regularization in the study 
of the contributions of vacuum energy effects at the cosmological 
scale. And we also summarize a number of important issues which 
are both undoubtedly beautiful (from the physical viewpoint) and 
useful in present-day observational cosmology. To finish, the 
reader should be warned that, for the reasons already given and 
lack of space, some fundamental issues, as inflation, quantum 
gravity and string theoretical fundamental approaches to cosmology 
will not be dealt with here. A minimal treatment of any of them 
would consume more pages than the ones at disposal and, again, a 
number of excellent treatments of these subjects are available. 
 
\end{abstract} 
 
\section{Introduction}    
 
Cosmology is the study of the world we live in. It would be 
difficult to find a more noble endeavour for a human being ---let 
aside perhaps from medicine--- given our proven capacity to 
`understand' things and events around us. To start, some already 
classic books, but very useful to everyone to get a basic 
knowledge of the cosmos are these in Ref.\cite{cosmol0} Here, 
let us just recall the several 
century long discussions, first on the issue of our Earth being 
flat or round, then about that same Earth being or not the center 
of the universe, and later the formulation of the 
universal law of gravitation  by Newton, who extended 
thereby the observation 
of the behavior of an apple falling from a tree on the earth 
surface to build a universal law for the whole of the cosmos. This 
immediately rises to the thinking mind (at least) two questions. 
\begin{itemlist} 
 \item  Why is the universe so `understandable' to the human 
mind? \item  And how  comes that mathematics are so useful in 
the formulation of the laws of nature? Just recall the extreme 
simplicity of Newton's formula, $F=GMm/r^2$, and the usefulness of 
Leibniz's calculus. Is this not very remarkable? 
\end{itemlist} 
 
Indeed, the question phrased by Eugene Wigner as that of {\it the 
unreasonable effectiveness of mathematics in the natural sciences} 
\cite{ew1} is an old and intriguing one. It goes back to 
Pythagoras and his school, ca. 550 BC ({\it ``all things are 
numbers''}), even probably to the sumerians, and maybe to more 
ancient cultures, which left no trace.  Immanuel Kant said that 
{\it ``the eternal mystery of the world is its 
comprehensibility''}\footnote{The reader should pause and take 
some minutes to meditate such a profound and mysterious statement. 
The author cannot refrain from confessing that he had arrived to 
the same conclusion by himself alone, before reading Kant. 
When he first 
discovered Kant's sentence, he was most deeply shocked.} and Albert 
Einstein contributed also to this idea in his 1936 essay ``Physics 
and Reality" where he elaborated on Kant's statement. Also {\it 
mathematical simplicity, and beauty,} have remained for many years 
crucial ingredients when having to choose among different 
plausible possibilities. 
 
Those are for sure profound and far reaching ideas by some of the 
people who established landmarks in the long way towards our 
present understanding of the Cosmos. However, one should be more 
humble and I would never 
 dare to use words like  {\it ``comprehend''} or 
 {\it `understand'}, but would rather replace them by {\it describe}, or 
{\it modelize}, in modern terms. Indeed, the fact e.g. that 
Newton's formula is so extremely simple and far reaching 
 does {\it not} mean at all that 
we {\it understand} the attraction of two massive bodies any 
better. Why do two bodies attract, and do not repel, each other? The 
only conclusion to be drawn is that Newton put us in possession of 
 a very simple, useful, accurate, and 
universal model for ordinary gravity. 
In trying to answer 
questions as the last one and others, such as the nature of the 
mass of a body, and in trying to formulate an ultimately universal 
law for the whole Universe  a lot of effort has been invested in 
the last 30 to 40 years. The results obtained have been 
extraordinary from the mathematical viewpoint (the new formalism 
itself and its mathematical applications, e.g. in algebraic 
geometry and knot theory). From the physical side the advance has 
been indeed consistent, but much less spectacular, and a unique 
{\it theory of everything} (TOE) is not in sight yet. 
 
In the last decade considerable effort has been put in 
observational cosmology with very rewarding results as, just 
 to mention two of them, the construction of the first maps of 
the cosmos (first including 
thousands and now millions of galaxies), and the discovery that 
the expansion of the universe we live in is accelerating. All 
indicates that this effort is going to continue, opening new 
perspectives for the awaited matching of the proposed mathematical 
models with the physical, observational results, 
 and also (not less important) for new jobs for cosmology students. 
 
These three lectures can only cover a very small part of what is 
known, or at least should be learnt, about cosmology. A personal 
bias is unavoidable and, far from trying to disguise it, the 
author's purpose is to complement the many excellent 
treatises that exists on the subject, by touching upon some 
matters of his competence that are not so often discussed, 
and trying also to bring 
together things that appear disconnected in the existing 
literature. The first lecture will be on {\it Mathematics as a 
tool to study the Universe}. Among the new mathematics required to 
study physical processes since the appearance of Quantum Field 
Theory (QFT),\cite{ramond1,bd1} there is the regularization or 
`summation' of infinite series. In the cosmological setting and in 
brane and string theory or Quantum Gravity,\cite{bos1} the 
most elegant method to do this has proven to be zeta function 
regularization, introduced by S. Hawking in 1975 with exactly this 
purpose.\cite{hawk1} A review of this 
technique\cite{kont95b,dzf1,zb1,zb2,zb3} will be presented in the 
first lecture, while a description of its uses in the calculation 
of the contribution of vacuum energy effects at the cosmological 
scale (e.g. in trying to explain the observed acceleration of the 
universe) will appear in the third. Also in the first lecture, an 
introduction of the formalism ordinarily used in Cosmology, as 
derived from the Einstein equations of General Relativity will be 
presented in some detail. 
 
The second lecture is devoted to some aspects of observational 
cosmology. We start by reviewing a number of important issues 
which are both extremely beautiful (from the physical viewpoint) 
and useful in present-day observational cosmology, such as the 
Hubble law, the Sunyaev-Zeldovich effect, the appearance of the 
Lyman-alpha forest structures, and the two variants of the 
Sachs-Wolfe effect. Then we present in some detail the 
mathematical techniques used for the purpose of the study of the 
large scale structure of our Universe. These include the analysis 
of point distributions (each point corresponding to a galaxy or 
galaxy cluster) and the use of non-standard statistics in the 
analysis of possible non-Gaussianities of the cosmic microwave 
background (CMB) temperature and matter fluctuations. Finally, as 
anticipated already, in the third lecture we will present some 
simple model in order to explain how quantum vacuum effects at 
cosmological scale can possibly contribute to the present value of 
the cosmological constant. For this purpose a further elaboration 
of the zeta regularization techniques, including some formulas 
originally derived by the author, will be necessary. 
 
There is no main conclusion to the paper, aside from the sum of 
the partial ones already suggested in the different sections. It 
is by now quite clear that the study of the Cosmos during the next 
decades will be most rewarding, not only intellectually or 
spiritually, but also in many more materialistic ways, for the 
many international collaborations already started or projected 
will undoubtedly require a large number of dedicated cosmologists 
in different tasks. Among them, to analyze and understand the 
observational results in the framework of existing and new 
ambitious theories, of which, for sure, there'll  still be 
plenty in the coming years.

\section{Mathematics as a tool to study the Universe} 
 
We now ellaborate on  the issues rised in 
the Introduction. 
 
\subsection{Some disgression on divergent series} 
 
The fact that the infinite series 
\begin{equation} 
s= \frac{1}{2} + \frac{1}{4} + \frac{1}{8} + \cdots 
\end{equation} 
has the value $s=1$ is nowadays clear to any school child. It was 
not so, in fact, for many centuries, as we can recall from {\it 
Zeno of Elea's paradox} (or Zeno's paradox of the tortoise and 
Achilles), transmitted by Aristotle and  based on the pretended 
impossibility to do an {\it infinite} number of summations (or 
recurrent `jumps' or steps of any kind, in a {\it finite} 
 amount of time). In fact there are still modern 
versions of the Zeno paradox (e.g. the quantum Zeno paradox) 
which pop up now and then.\cite{qzp1} 
We shall not discuss those here, but rather assume that the 
following process is clear to the reader: by taking the first 
term, $1/2$, to the left what remains on the r.h.s. 
 is just one half of the 
original series (i.e., $1/2$ is a common factor), 
so that 
\begin{equation} 
s- \frac{1}{2} = \frac{s}{2}    \ \ \ \ \Longrightarrow \ \ \ \ s=1. 
\end{equation} 
Now, what is the sum of the following series? 
\begin{equation} 
s=1+1+1+1+\cdots 
\end{equation} 
Again, any school child would answer immediately that $s= \infty$. 
In fact, whatever $\infty$ may be, everybody recognizes in this 
last expression the definition itself of the infinity: the piling 
of one and the same  object, once and again, without end. 
Of course, this is true, 
but it is useless to modern Physics,  being precise, since the 
advent of quantum fields. Calculations there are plagued with 
divergent series, and it is of no use to say that this series is 
divergent, and that other one is also divergent, and the other 
there too, and so on. One gets non-false but useless information 
in this way, since we {\it do not see} these infinites in Nature. 
 
For years, there was the suspicion that one could try to give 
sense to divergent series. This has proven (experimentally!) to be 
true in Physics, but it were the mathematicians ---many years 
before--- who first realized this possibility. In fact, Leonard 
Euler (1707-1783) was convinced that {\it ``To every series one 
could assign a number''}\cite{euler1} (that is, in a reasonable, 
consistent, and useful way, of course).  Euler was unable to prove 
this statement in full, but he devised a technique (Euler's 
summation criterion) in order to `sum' large families of divergent 
series. His statement was however controverted by some other great 
mathematicians, as Abel, who said that {\it ``The divergent series 
are the invention of the devil, and it is a shame to base on them 
any demonstration whatsoever"}.\cite{abel1} There is a classical 
treatise due to G.H. Hardy and entitled simply {\it Divergent 
series}\cite{hardy1} that I highly recommend to the reader. 
 
As always with modern Mathematics, one starts the attack on 
divergent series by invoking a number of axioms, like 
\begin{romanlist}[(ii)] 
\item If $a_0+a_1+a_2+\cdots =s$, then $ka_0+ka_1+ka_2+\cdots 
=ks$. \item If $a_0+a_1+a_2+\cdots =s$, and $b_0+b_1+b_2+\cdots 
=t$, then \\ $(a_0+b_0)+(a_1+b_1)+(a_2+b_2)+\cdots =s+t$. \item If 
$a_0+a_1+a_2+\cdots =s$, then $a_1+a_2+\cdots =s-a_0$. 
\end{romanlist} 
A couple of examples.\begin{romanlist}[(b)] \item Using the third 
axiom we obtain that for the series $s=1-1+1-1+\cdots$, we have 
$s=1-s$, and therefore $s=1/2$. This value is easy to justify, 
since the series is oscillating between 0 and 1, so that 1/2 is 
the more `democratic' value for it. \item Using now the second 
axiom, we obtain that for the series $t=1-2+3-4+\cdots$, it turns 
out by subtracting it term by term from the former one that 
$s-t=t$, and therefore $t=s/2=1/4$. Such result is already quite 
difficult to swallow. This is in common with most of the finites 
values that are obtained for infinite, divergent series. 
\end{romanlist} 
But, what to do about our initial series $1+1+1+\cdots$? This one is 
most difficult to tame, and the given axioms do not serve to 
this purpose. But there is more to the axioms, which are only 
intended as a humble starting point. By reading Hardy's book one 
learns about a number of different methods that have been proposed 
and is good to know. They are due to Abel, Euler, Ces{\`a}ro, 
Bernoulli, Dirichlet, Borel and some other 
mathematicians.\footnote{Pad\'e approximants should 
in no way  be forgotten in this discussion.\cite{pade1}}  The 
most powerful of them involve analytical continuation in the 
complex plane, as is the case of the so called zeta regularization 
method. 
 
Thus, for instance, a series 
\begin{equation} 
a_0+a_1+a_2+ \cdots 
\end{equation} 
 will be said to be {\it Ces{\`a}ro summable}, and its sum to be the number $s$, 
if the limit of partial sum means exists and gives $s$, namely 
\begin{equation} 
\exists \, \lim_{n \to \infty} \sum_{n=1}^\infty \frac{A_n}{n} = 
s, \qquad  A_n\equiv \sum_{j=1}^n a_j. 
\end{equation} 
This criterion can be extended and gives rise to a whole family of 
criteria for {\it Ces{\`a}ro summability}.  On the other hand, the 
series before will be said to be {\it Abel summable}, and its sum 
to be the number $s$, 
\begin{equation} 
\sum_{n=0}^\infty a_n =s, 
\end{equation} 
 if the following function constructed as a power series 
 \begin{equation} 
f(x) \equiv \sum_{n=0}^\infty a_n x^n 
\end{equation} 
is well defined for  $0<x<1$ and the limit when $x$ goes to 1 from 
the left exists and gives $s$, namely, 
\begin{equation} 
\exists \, \lim_{x\to 1^-} f(x) =s. 
\end{equation} 
And similarly for the rest of the criteria, which are not 
equivalent, as one can check.\cite{hardy1} 
 
\subsection{Zeta regularization in a nutshell} 
As advanced already, the regularization and renormalization 
procedures are essential issues of contemporary physics ---without 
which it would simply not exist, at least in the form we know 
it.\cite{renorm1} 
 Among the different methods, zeta function regularization 
---which is obtained by analytical continuation  in the complex plane 
of the zeta function of the relevant physical operator in each 
case--- is maybe the most beautiful of all. Use of this method 
yields, for instance, the vacuum energy corresponding to a quantum 
physical system (with constraints of any kind, in principle). 
Assume the corresponding Hamiltonian operator, $H$, has a 
spectral decomposition of the form (think, as  simplest case, in a 
quantum harmonic oscillator): $\{ \lambda_i, \varphi_i \}_{ i\in 
I}$, being $I$ some set of indices (which can be discrete, 
continuous, mixed, multiple, \ldots). Then, the quantum vacuum 
energy is obtained as follows:\cite{zb1} \begin{equation} 
\sum_{i\in I}\left( \varphi_i, H  \varphi_i \right) = 
\mbox{Tr}_{\zeta} 
H = \sum_{i\in I} \lambda_i = \left.  \sum_{i\in I} \lambda_i^{-s} 
\right|_{s=-1} = \zeta_H (-1), \end{equation} 
 where $\zeta_H$ is the zeta 
function corresponding to the operator $H$, and the equalities are 
in the sense of analytic continuation (since, generically, the 
Hamiltonian operator will not be of the trace class).\footnote{The 
reader should be warned that this $\zeta-$trace is actually 
no trace in the usual sense. In particular, it is highly non-linear, 
as often explained by the author elsewhere. Some colleagues are 
unaware of this 
 fact, which has lead to important mistakes and erroneous conclusions 
too often.} 
 Note that the formal 
sum over the eigenvalues is usually ill defined, and that the last 
step involves analytic continuation, inherent with the definition 
of the zeta function itself. 
 
The method evolved from the consideration of the Riemann zeta 
function. This was introduced by Euler, from considerations of the 
harmonic series 
\begin{equation} 
1+\frac{1}{2} + \frac{1}{3} + \frac{1}{4} + \cdots, 
\end{equation} 
which is logaritmically divergent, and of the fact that puttin a 
real exponent $s$ over each term 
\begin{equation} 
1+\frac{1}{2^s} + \frac{1}{3^s} + \frac{1}{4^s} + \cdots, 
\end{equation} 
then for $s>1$ the series is convergent, while for $s\leq 1$ it is 
divergent. Euler called this expression, as a function of $s$, the 
$\zeta-$function, $\zeta (s)$, and found the following important 
relation 
\begin{equation} 
\zeta (s) = \sum_{n=1}^\infty \frac{1}{n^s} = \prod_{p\, 
\mbox{\footnotesize prime}} \left( 1- \frac{1}{p^s} \right)^{-1}, 
\end{equation} 
which is crucial for the applications of this function in Number 
Theory. By allowing the variable $s$ to be complex, Riemann saw 
the relevance of this function (that now bears his name) for the 
proof of the prime number theorem\footnote{Which states that the 
number $\Pi (x)$ of primes which are less or equal to a given 
natural number $x$ behaves as $x/\log x$, as $x \to \infty$. It 
was finally proven, using Riemann's work, by Hadamard and de la 
Valle Poussin.}, and formulated thereby the {\it Riemann 
hypothesis}, which is one of the most important problems (if not 
{\it the} most) in the history of Mathematics. More to that in the 
excellent review by Gelbart and Miller.\cite{gelmi1} 
 
\begin{figure} 
\centerline{\psfig{file=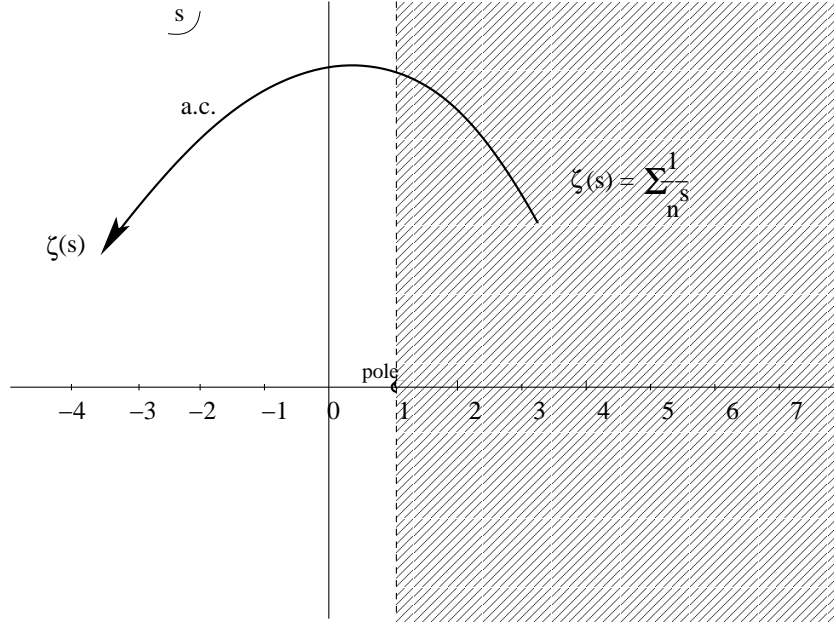,width=11cm}} 
\vspace*{8pt} 
\caption{The zeta function $\zeta (s)$ is defined in the following 
way, 
on the whole complex plane, $s\in$ {\bf C}. To start, on the open 
half of the complex plane which is on the r.h.s of the abscissa of 
convergence Re$\,s =1$, $\zeta$ is defined as the absolutely 
convergent series: $\zeta (s) = \sum_{n=1}^{\infty} n^{-s}$. In the 
rest of the $s-$complex plane, $\zeta (s)$ is defined as the (unique) 
analytic continuation of the preceding function, which turns out to be 
meromorphic. Specifically, it is analytic everywhere on the complex 
plane except for one simple pole with residue equal to 1, 
which is at the point $s=1$ (notice that it corresponds to the 
logarithmically divergent harmonic series, as already discussed).} 
\label{zetaf1} 
\end{figure}

For the Riemann $\zeta (s)$, the corresponding (complex) series 
converges absolutely for the (open) half of the complex plane to 
the right of the abscissa of convergence Re $s =1$, while it 
diverges on the other side, but it turns out that it can be 
analytically continued to that part of the plane, being then 
everywhere analytic (and finite) except for the only, simple pole 
at $s=1$ (Fig. \ref{zetaf1}).\footnote{Where it yields the 
harmonic series: there is 
no way out for this one.} 
In more general cases, namely corresponding to the Hamiltonians which are 
relevant in physical applications,\cite{zb1,zb2,zb3} the situation is 
in esence quite similar, albeit in practice it can be rather more 
involved. A mathematical theorem exists, which assures that under 
very general conditions the zeta function corresponding to a Hamiltonian 
operator will be also meromorphic, with just a discrete number of possible 
poles, which are simple and extend to the negative side of the real 
axis.\footnote{Although there are some exceptions to this general behavior, 
they correspond to rather twisted situations, and are outside 
the scope  of this brief presentation.} 
 
The above picture  already hints towards the use of 
the zeta function as a summation method. Let us consider two 
examples. 
\begin{romanlist} \item We interprete our starting series 
\begin{equation} 
s_1=1+1+1+1+\cdots 
\end{equation} 
as a particular case of the Riemann zeta function, e.g. for the 
value $s=0$. This value is on the left hand side of the abscissa 
of convergence (Fig. \ref{zetaf1}), where the series as such 
diverges but where the 
analytic continuation of the zeta function provides a perfectly 
finite value: 
\begin{equation} 
s_1=\zeta (0) =-\frac{1}{2}. 
\end{equation} 
So this is the value to be attributed to the series 
$1+1+1+1+\cdots$. \item The series 
\begin{equation} 
s_2=1+2+3+4+\cdots 
\end{equation} 
corresponds to the exponent $s=-1$, so that 
\begin{equation} 
s_2=\zeta (-1) =-\frac{1}{12}. 
\end{equation} 
\end{romanlist} 
A couple of comments are in order. 
\begin{itemlist} 
 \item In a short period of less than a year, two distinguished 
 physicists, A. Slavnov and F. Yndurain, gave seminars in Barcelona, 
about different subjects. It was remarkable that, in both 
presentations, at some point the speaker addressed the audience 
with these words: 
{\it ``As everybody knows, $1+1+1+ \cdots 
=-1/2$''}.\footnote{Implying maybe: {\it If you do not know this it 
is no use to continue listening.} Remember by the way the {\it 
lemma} of the Pythagorean school: {\it Do not cross this door if 
you do not know Geometry.}} 
 \item That positive series, as the ones above, can yield a 
 {\it negative} result may seem uttermost nonsensical. However, it turns 
 out that the most precise experiments ever carried out in Physics do confirm 
such results. More exactly:  models of regularization in QED built 
upon these techniques lead to final numbers which are in 
agreement with the experimental values up to the the 14th 
figure.\cite{g21} In recent experimental proofs of the Casimir 
effect\cite{Casimir} the agreement is also quite remarkable (given 
the difficulties of the experimental setup).\cite{casexp1} 
\item The method of zeta regularization is based on the analytic continuation 
of the zeta function in the complex plane. Now, how easy is to perform 
 that? Will we need to undertake a lengthy complex plane computation every 
time? It turns out that this is not so. The result is immediate, in 
principle, once you know the appropriate reflection 
formula (also called functional equation) that your zeta function obeys. 
In the case of the Riemann zeta: 
\begin{equation} 
\xi (s) =\xi (1-s),   \qquad  \xi (s) \equiv \pi^{-s/2} \Gamma 
(s/2) \zeta (s). 
\end{equation} 
 In practice these formulas are however not optimal for actual calculations, 
since they are ordinarily given in terms of power series 
expansions (as the Riemann zeta itself!). Fortunately, sometimes 
there are more clever expressions, that can be found, which converge 
exponentially fast,  as the Chowla-Selberg\cite{cs1} formula and 
some others.\cite{eecs1} Those give real power to the method of zeta 
regularization. More about this point in lesson 3, where a number 
of such 
expressions will be explicitly used in the calculation of the 
contribution of the quantum vacuum fluctuations to the 
cosmological constant. 
 
\end{itemlist}

\section{Observational Cosmology: large scale structure}

\subsection{A landmark in observational cosmology} 
 
Redshift surveys of galaxies, being three-dimensional, do not 
suffer from projection effects of two-dimensional maps on the sky 
surface and are much more appropriate to obtain the true large 
scale structure of our Universe. However, not all the contribution 
to the redshift comes from the cosmological expansion 
(which defines the third dimension, along the line of sight), since there 
are also additional contributions coming from the peculiar velocities 
of the galaxy in question (attraction of other galaxies in a cluster, 
displacement of the cluster itself, etc.):\footnote{This formula is just an 
approximation, since as space is curved, for non-near galaxies the 
distance/redshift relation is non-linear.} 
\begin{equation} 
cz_{observed} =cz_{cosmological} +v_{peculiar}. 
\end{equation} 
This can originate artifacts such as the ``finger of God" effect 
in which clusters of galaxies appear as long fingers pointing 
radially towards the observer. It is not easy to correct for these 
effects, so one must be careful when trying to make sense of such 
structures and of three-dimensional maps in general. 
 
\begin{figure}[!ht] \vspace*{-30pt} 
\centerline{\psfig{file=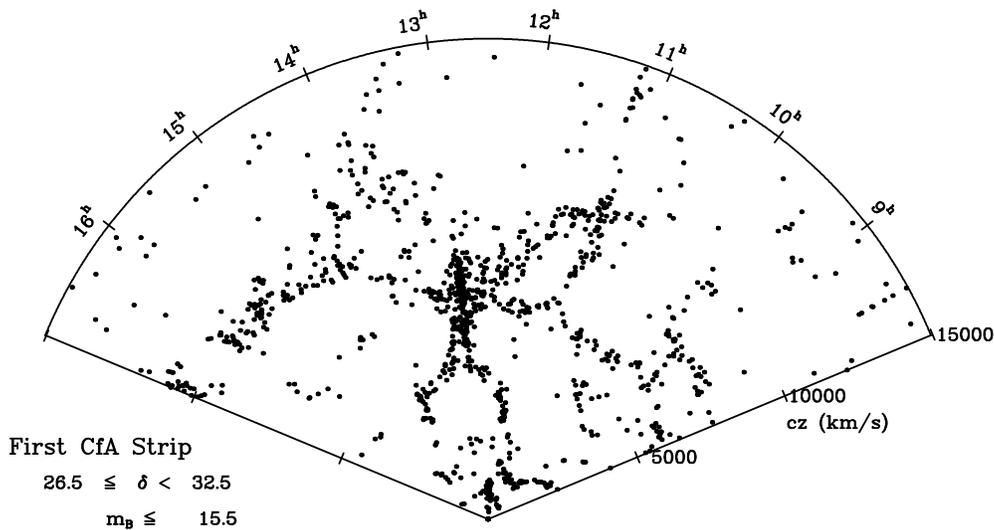,angle = -90, width=14cm}} 
\vspace*{-30pt} \caption{The first CfA redshift survey caused an 
immediate impact on the scientific community. It clearly showed 
that the distribution of galaxies in space was anything but 
random, with galaxies actually appearing to be distributed on 
surfaces, almost bubble like, surrounding large empty regions, or 
``voids.'' V. de Lapparent, M. Geller and J.P. Huchra, Astrophys. 
J. Lett. {\bf 302}, L1 (1986) Smithsonian Astrophysical 
Observatory.} \label{cfa1} 
\end{figure}

The CfA redshift survey of de Lapparent, Geller and Huchra  (1986, 
1988) was a landmark.\cite{dlgh1} It was just {\it ``A slice of 
the Universe''}, but for the first time, in a map of black dots 
where each dot corresponded to a whole galaxy,  the large scale 
structure of our Universe (a map of our world) appeared in front 
of our eyes, for the very first time ever (Fig. \ref{cfa1}). 
 That survey, and the 
corresponding Southern Sky Redshift Survey (da Costa et al., 
1988),\cite{daco1} showed the by now familiar filaments and walls 
surrounding  voids:\cite{jones1} the ``bubble-like" textures of 
the galaxy distribution,  on scales where the galaxy-galaxy 
correlation function is negligible. The impact was immediate, and 
influenced a large amount of physicists working in different 
subjects, who tried to explain, modelize and even reconstruct the 
point distribution of the slice, in terms of more or less 
fundamental theories.

A paper by J. Ostriker, C. Thomson, and E. 
Witten\cite{otw1} tried to explain the voids and other structures 
as a consequence of string theory. I tried to address the much 
more technical (but in my view not less important) issue of how to 
perform the comparison of two point distributions. They would be, 
in the case under study,  the observed galaxy distribution for the 
slice geometry and any  point distribution obtained, say, from a 
simulation of a theoretical model which would pretend to yield 
`the same' or a good  approximation to the observed point map. The 
simple (but very difficult) question to be answered is just: how 
close are the observed map and the one obtained from a model? Of 
course, a rigorous answer can be given, from Statistics, 
in terms of the 2-point, 
3-point, ..., $n-$point correlation functions of both point 
distributions. But it turns out that in practice higher order 
correlation functions are very difficult to compute for a large 
sample of points, and one has to find alternative, much more 
direct and optimized routes: the highest possible 
 unbiased information from 
the lowest number of moments of the distribution. One of methods 
we considered are the so-called {\it counts in cells}. I suggested 
this as the starting point for a PhD Thesis to Enrique 
Gazta{\~n}aga,  who had come to me at the appropriate time in 
search of a subject to work on. We did quite nice work together on 
these matters, later extended by Pablo Fosalba and Jose 
Barriga with considerable success.\footnote{This was  the seed 
and the beginning of our 
cosmology group at the IEEC/CSIC Institute in Barcelona. Presently 
we are involved in PLANCK's\cite{planck1} High and  Low 
Frequency Instruments,\cite{hfilfi} the Sloan SDSS,\cite{sdss} the 
APM,\cite{apm} the 2dF,\cite{2df} WMAP, 
\cite{wmap} etc.}

\begin{figure}[!th] 
\centerline{\psfig{file=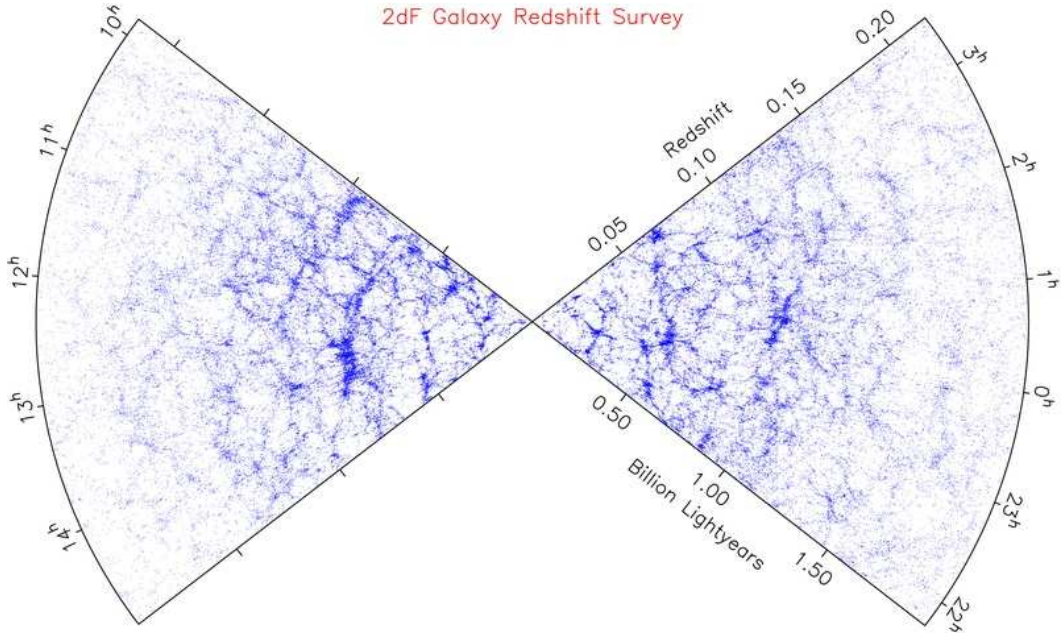, width=14cm}} 
\vspace*{8pt} 
\caption{Final data release, of 30 June 2003, of the 2dF Galaxy 
Redshift Survey (Matthew Colless, Steve Maddox,  John Peacock, et al., 
Anglo-Australian Observatory,
http://magnum.anu.edu.au/$\sim$TDFgg/). The figure shows the map of the 
galaxy distribution produced from the completed survey.} \label{2df} 
\end{figure}

Two important redshift surveys have been based on the IRAS 
catalogue (the IRAS survey and QDOT). 
 The original surveys have now been extended to other 
slices and the original few thousands of galaxies of CfA are now 
transformed into the several millions monitored by the 2 Degree Field 
survey, 2dF, the Sloan Digital Sky Survey, SDSS, etc. 
The most recent results of the 2dF survey are depicted in Fig. \ref{2df}.
Comparison with Fig. \ref{cfa1} of the first CfA redshift survey shows 
the enormous progress in observational cosmology in the last 15 years.
The observations 
 seem to conclude that the structures of the large scale galaxy 
point distribution are essentially sheet-like, and that the scale 
of the sheets is limited only by the scale of the survey. The most 
remarkable feature, the so-called ``great wall" (Geller and 
Huchra, 1989)\cite{gh1} has been seen to be enhanced by the 
selection function for the sample, but it also appears in other 
deep wide angle surveys. In any case, there has been much 
discussion about this issue. Great walls do not bound great voids, 
but seem to surround collections of smaller voids that are 
themselves bounded by not-so-great walls. It could be that the 
great walls are  picked out and correlated by our own eyes to 
build a larger structure.\cite{lsst1}  Looking at N-body models 
suggests this kind of effect because it is easy for the brain to 
identify coherent structures on scales where there is no physical 
mechanism for generating structure.\cite{jones1} 

Another very important survey is 
the APM Galaxy Survey, which contains positions, magnitudes, sizes and 
shapes for about three million galaxies selected from 269 UKST 
survey plates which were scanned using the APM Facility. 
The galaxies have apparent magnitudes with $17 < b_j < 20.5$ and are 
spread through the largest volume of the universe that has been surveyed 
to date. The picture to be seen in the  APM Galaxy Survey page 
(http://www.nottingham.ac.uk/$\sim$ppzsjm/apm/apm.html)
 shows the galaxy distribution as a 
density map in equal area projection on the sky, centered on the South 
Galactic pole. Each pixel covers a small patch of sky 0.1 
degrees on a side, and is shaded according to the number of galaxies 
within the area: where there are more galaxies, the pixels are 
brighter. Galaxy clusters, containing hundreds of galaxies closely 
packed together, are seen as small bright patches. The larger 
elongated bright areas are superclusters and filaments. These 
surround darker voids where there are fewer galaxies. The colours
 are  coded according to the apparent magnitude of the galaxies in 
each pixel: fainter galaxies are shown as red, intermediate are shown 
as green and bright are show as blue. The more distant galaxies tend to 
be fainter, and also show less clustering, and so the maps 
has a generally uniform reddish background. The more nearby galaxies 
tend to be bright, and are more clustered, so the more 
prominent clusters of galaxies in the map tend to show up as blue. The 
small empty patches in the map are regions that we have 
excluded around bright stars, nearby dwarf galaxies, globular clusters 
and step wedges. It is very adviceable for the reader to look at this picture
in http://www.nottingham.ac.uk/$\sim$ppzsjm/apm/apm.html, by Steve Maddox, 
Will Sutherland,  George Efstathiou and Jon Loveday. 
 
\subsection{The two pillars of observational cosmology, and other tools} 
 
The two major pillars of observational cosmology are, undoubtedly: 
\begin{itemlist} 
 \item the CMB temperature fluctuations, and 
 \item the fluctuations of the matter distribution (of galaxies), 
or density fluctuations. 
\end{itemlist} 
From a technical viewpoint, observation gained an enormous thrust 
with the advent of multi-fiber spectrographs, with extremely 
complex astronomical instrument (as that of the 2dF): from the 5 to 10 
redshifts per night that were produced with the old methods not 
so long ago one 
has got to the 100 to 2000 that can be taken now under good 
observation conditions. 
 
\subsubsection{Density fluctuations} 
 
The density fluctuations, $\delta$, (generated e.g. by inflation) 
in a homogeneous universe of mean density $\bar{\rho}$, are such 
that:\cite{gaztalec1} 
\begin{equation} 
\rho (x) = \bar{\rho} \left[ 1+ \delta  (x) \right], \qquad \delta 
= \frac{\rho -\bar{\rho}}{\bar{\rho}}, 
\end{equation} 
where $\rho (x)$ is the density around a given point, $x$, and 
$\delta (x)$ the corresponding fluctuation around it. 
In the fluid limit (non-crossing orbits) the evolution of 
fluctuations is then governed by the basic laws: 
\begin{itemlist} 
 \item the  Friedmann equation, 
\item the continuity equation, and 
\item the Poisson equation, 
\end{itemlist} 
 which in Fourier 
space and in terms of the corresponding Fourier decomposition of 
the density fluctuations, namely, 
\begin{equation} 
\delta  (x) = \sum_k \delta_k e^{-ikx}, 
\end{equation} 
result in the following fundamental equation: 
\begin{equation} 
\frac{d^2\delta_k}{d\eta^2} + {\mathcal H} 
(\eta)\frac{d\delta_k}{d\eta} - \left[ \frac{3}{2} {\mathcal H} 
(\eta) \Omega_M (\eta) - k^2 v_s^2 \right] \delta_k =0, 
\end{equation} 
where $v_s$ is the sound velocity in the fluid ($v_s^2= \partial p 
/\partial \rho$), $\eta$ the conformal time ($d\eta= dt/a$), and 
\begin{equation} 
{\mathcal H} \equiv \frac{d \ln a}{d\eta} = a H, \quad \Omega_M 
(\eta) \equiv \frac{\rho_M (\eta)}{\rho_c} = \frac{\Omega_M 
a^{-3}}{H^2}, 
\end{equation} 
with $H$ the Hubble constant at present, and similarly  $\Omega_M$. 
 
There are two different regimes encompasing 
 the solutions of the above equation, 
namely: 
\begin{romanlist}[(ii)] 
\item  growing fluctuations, when the term in brackets is 
positive, and 
\item amortiguation, when it is negative, separated 
by the 
\item Jeans scale $\lambda_J$ , obtained when the term in brackets 
vanishes, that is, for 
\begin{equation} 
k_J = \frac{2\pi}{\lambda_J}, \quad  k_J^2 = \frac{3}{2} 
\frac{{\mathcal H}^2\Omega_M}{v_s^2}. 
\end{equation} 
\end{romanlist} 
 
\subsubsection{Power spectrum} 
 
The power spectrum is the mean quadratic value of the fluctuation 
amplitude $k$ mode, that is 
\begin{equation} 
P(k) = < \delta_k^2 >. 
\end{equation} 
Owing to the fact that quantum fluctuations in the fundamental 
state are scale invariant, it turns out that the primordial 
spectrum, which is believed to have been originated by quantum 
fluctuations, must be scale invariant, too. This translates into a 
density spectrum of the kind: 
\begin{equation} 
P_0(k) = < \delta_k (0)^2 > = A \, k^{n_s}, \quad n_s \simeq 1, 
\end{equation} 
which is the Harrison-Zeldovich spectrum. 
 
The Jeans instability produces a definite break of this behavior, 
at about 
\begin{equation} 
k_{break} \simeq 0.05 \, h /\mbox{Mpc}, \quad \lambda_{break} 
\simeq 60 \, \mbox{Mpc}/h, 
\end{equation} 
which is when matter domination starts to take over. Because of the 
uncertainty about the exact value of the Hubble constant (that is 
something around $100$ km/s Mpc, it is very common 
to write this uncertainty as a coefficient, $h$, so that the true value is 
$H = 100\, h$ km/s Mpc, and $h$ is a number somewhere between 
0.6 and 0.8. This $h$ appears in many distance estimates, as in 
the expressions above. 
 
\subsubsection{Temperature spectrum} 
 
A similar analysis can be repeated for the temperature 
fluctuations of the CMB. However, the CMB is seen by us as a 
two-dimensional projection of the surface of last scattering, that is the 
moment, when the universe was about 300.000 years old, in that 
it was cool enough in order to suddenly become transparent to radiation. 
For this reason, it turns out that 
the Fourier decompositions is here a spherical 
harmonics decomposition, so that: 
\begin{equation} 
\delta_T (\theta, \varphi) = \sum_{l=0}^\infty 
\sum_{m=-\ell}^{+\ell} a_{\ell m} Y_{\ell m} (\theta, \varphi), 
\quad \ell = \frac{\pi}{\theta}, 
\end{equation} 
and the temperature power spectrum is given in terms of the coefficients 
\begin{equation} 
c_\ell = < a_{\ell m}^2 > = \frac{1}{2\ell +1} 
\sum_{m=-\ell}^{+\ell} |a_{\ell m}|^2, 
\end{equation} 
where the fair-sample hypothesis is used (namely, the mean 
over realizations equals the spatial mean). 
 
\begin{figure}[!ht] 
\centerline{\psfig{file=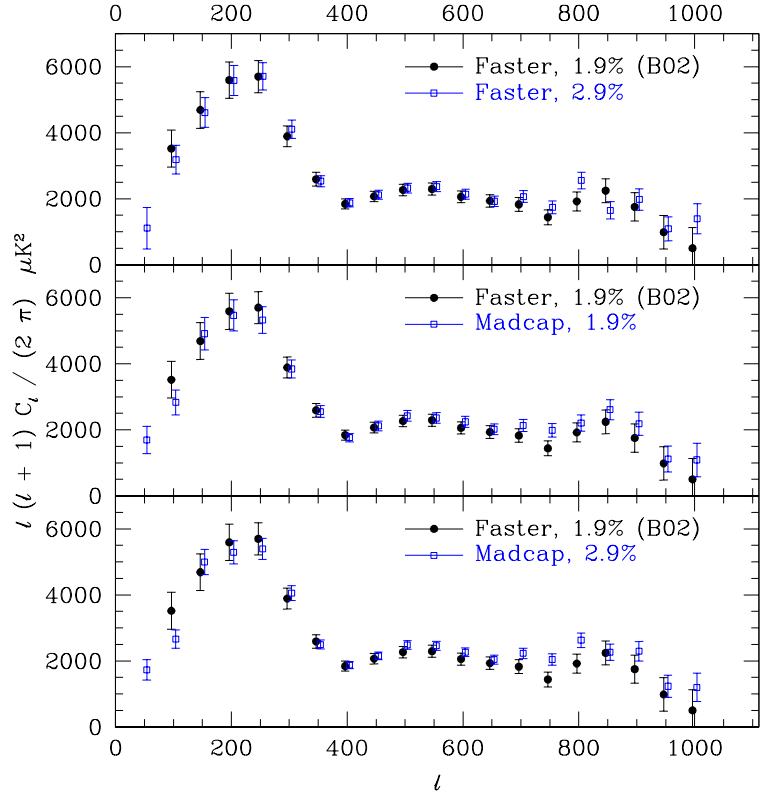,width=11cm}} 
\vspace*{1pt} 
\caption{A comparison of two recent analyses of 
observations with BOOMERanG. 
The agreement is generally very good, with the greatest variations at 
high $\ell$ where noise, rather than cosmic variance, dominates the 
errors.  J.E. Ruhl, et al., Astrophys. J. {\bf 599}, 786 (2003), 
Fig. 7.} \label{psboom1} 
\end{figure}

\begin{figure}[!ht] 
\centerline{\psfig{file=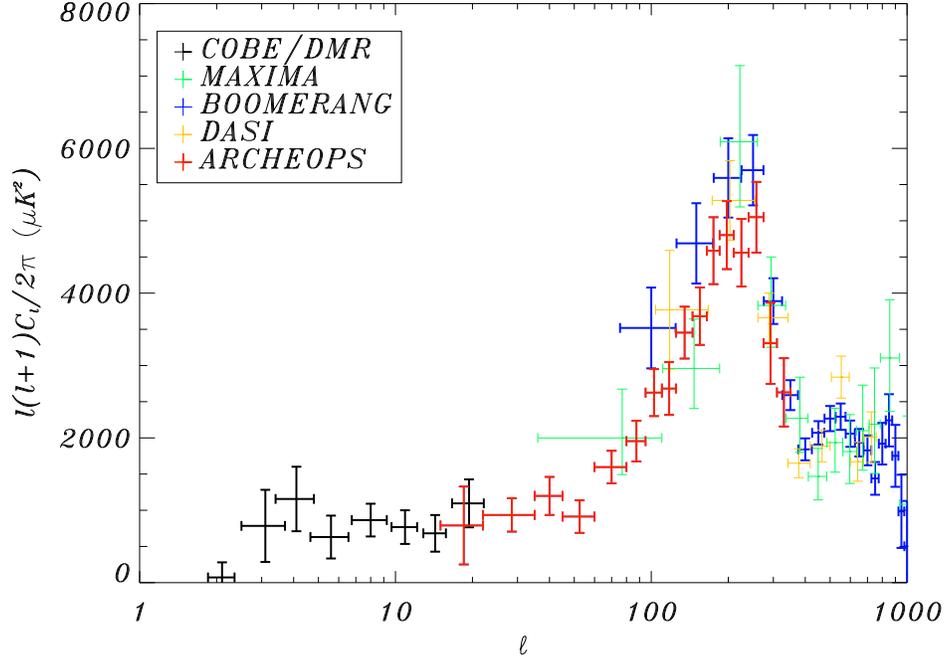,width=13cm}} 
\vspace*{1pt} 
\caption{The Archeops power spectrum compared with 
results of COBE, BOOMERANG, DASI, and MAXIMA). 
A. Benoit, et al., Astron. Astrophys. {\bf 399}, L19 
 (2003), Fig. 4.} \label{pswmap1} 
\end{figure}

The temperature power spectrum is the most basic reference tool used 
nowadays for the comparison of different models, in particular, the 
adjustment of the different peaks of the spectrum (the first being 
the acoustic peak). In the two following figures we see, respectively, 
an improved measurement of the angular power spectrum of temperature 
anisotropy in the CMB obtained from two recent analyses of 
observations with BOOMERanG (Fig. \ref{psboom1}), 
and the one 
measured by Archeops  (Fig. \ref{pswmap1}), which is conceived 
as a precursor of the PLANCK HFI instrument, 
 using the same optical design and technology. 

\subsubsection{The two-point correlation function} 
 
Another tool impossible to dismiss was the 
definition of the two-point correlation function, by Totsuji and 
Kihara in 1969,\cite{tk1} as the excess function, $\xi (r)$, appearing in 
the following expression for the number of galaxies to be found in 
a volume $dV$ around the position $r$: 
\begin{equation} 
dP= n\left[ 1+ \xi (r) \right] dV, 
\end{equation} 
$n$ being the average galaxy number density. The following scaling 
law is basic: the number galaxies (points) to be found in a sphere 
of radius $a$ behaves as a certain power of $a$, for  $a$ large 
enough, namely 
\begin{eqnarray} 
N_{<a} &=& n \int_0^a 4\pi r^2 \left[ 1+ \xi (r) \right]\,  dr 
\nonumber \\ & \propto&  a^{d_2}. 
\end{eqnarray} 
In sky surveys, the value of $d_2$ has been seen to evolve from 
$d_2=2$ to $d_2=3$, when one goes from intermediate to larger 
(that is, $a > 30 h^{-1}$ Mpc) scales. 
 
Now, recall the behavior of the energy density, for large distance 
$a$, as a function of $a$, for a matter dominated and a radiation 
dominated universe, namely 
\begin{equation} 
\rho (a) \sim \rho_0 \, a^{-s} \qquad \left\{ \begin{array}{l} s=3, \ 
\ \mbox{matter dominated,} \\ 
s=4, \ \ \mbox{radiation dominated.} \end{array} \right. 
\end{equation} 
And from Friedmann's equation 
\begin{equation} 
\frac{\dot{a}}{a} = \frac{8}{3} \pi G \rho_0 a^{-s} - 
\frac{k}{a^2}, \end{equation} by taking an additional derivative, 
we obtain 
\begin{equation} 
\frac{\ddot{a}}{a} = -(s-2) \frac{4}{3} \pi G \rho (a), 
\end{equation} 
which (remarkably) it is independent of $k$, and tells us at once 
that the expansion of the universe is decelerated by gravity. 
Also, that since we do not know of any form of energy with a 
behavior corresponding to  a power $s<2$, let aside of course from the 
cosmological constant, which has $s=0$: 
\begin{equation} 
\rho_{\Lambda} =\frac{\Lambda}{8\pi G}, 
\end{equation} 
 it turns out that the only possible causes of 
accelerated expansion can be the said cosmological constant or 
either a (still unknown) form o exotic matter or energy. In the 
case of the cosmological constant, $\Lambda$, it could come from 
$T_{\mu\nu}$ as a sort of vacuum contribution, so that 
\begin{equation} 
\rho (a) = \rho_M^{now} a^{-3} + \rho_R^{now} a^{-4} + 
\rho_{\Lambda}, 
\end{equation} 
 where {\it now} referes to $a=1$. Two kind of different solutions 
to this puzzle have been proposed: 
\begin{romanlist}[(ii)] 
\item to modify the curvature or 
geometric part of Einstein's equations with the addition of terms 
of the sort $1/R^2$, ln $R$, or other convenient functions of the 
curvature, $f(R)$,\cite{fR1} which sometimes produced the 
so-called phantom 
matter (sometimes giving rise to a new, future singularity, the 
big RIP, see, e.g., Ref.\cite{eno04a} and the references therein) or, 
  alternatively, 
\item to modify the matter part of the 
equations with the addition of true exotic matter or energy (see, 
e.g., Ref.\cite{gasp1} and the references therein). 
\end{romanlist} 
 
\subsection{A summary of cosmological facts and important effects} 
 
\subsubsection{The Olbers' paradox (1757-1840)} Should be known 
to everybody but, just in case ... By looking at the sky, Olbers 
came to the thought that if the Universe were infinitely old and 
infinite in extent and stars could shine forever, then every 
direction you looked into would eventually end on the surface of a 
star: the whole sky should be as bright as the surface of the Sun. 
 
Absorption by interstellar dust does not circumvent this paradox, 
since dust re-radiates whatever radiation it absorbs within a few 
minutes. 
 
{\it Solution.} The Universe is {\it not} infinitely old and its 
{\it expansion} reduces the accumulated energy radiated by distant 
stars. Either one of these effects acting alone already solves the 
paradox. 
 
\subsubsection{The Sunyaev-Zeldovich effect} 
 
Hot gas in clusters of galaxies distorts the spectrum of the CMB 
radiation.  Hot electrons there scatter a small fraction of the 
CMB photons and replace them with slightly higher energy photons. 
That is, high energy  electrons in the clusters and the 
homogeneity and isotropy of the CMB results in the CMB  photons 
gaining energy by Compton scattering.\cite{sz1} 
 The difference between the CMB seen through the cluster 
and the unmodified CMB can be measured. 
 
The effect, first described by R. Sunyaev and Ya.B. Zeldovich, 
 is observed as a  deficit of about 
$0.05\%$  of CMBR photons, as the ``missing" photons have been 
shifted to higher energy, with an increase of about $2\%$. Thus, 
the CMB radiation denounces the presence  of galaxy clusters found 
in its way towards us. 
 
This effect verifies the cosmological origin of the CMB. Moreover, 
combining radio with X-ray 
observations of cluster allows to determine: 
\begin{itemlist} 
 \item  the distance of the cluster; 
 \item  the value of the Hubble constant $H_0$; and 
\item for very distant clusters,  the value of the deceleration parameter. 
\end{itemlist}

\subsubsection{The Sachs-Wolfe effect} 
 
Photons are subject to the influence of gravity (a GR effect). When passing 
through a higher (resp. lower) concentration of matter, they  undergo a 
redshift (resp. a blueshift). 
R.K. Sachs and A.M. Wolfe where the first to realize this sort of effect 
should take place, under the 
form of perturbations of a cosmological model and 
angular variations of the cosmic microwave background. They also 
described an integrated form of the effect. 
 
{\it The integrated Sachs-Wolfe effect.} 
Consists on the gravitational redshift induced by photons falling into and 
climbing out of regions of space with different matter density (potential 
wells), between the Earth and the surface of last scattering. In contrast, the 
non-integrated Sachs-Wolfe effect is only at the surface of last 
scattering itself. 
 
\subsubsection{The Lyman Alpha forest} 
 
Some spectacular pictures of the universe show that 
we actually live in a forest, made of ``trees" of hydrogen gas 
which absorb light from 
distant objects. It leaves numerous absorption lines in a distant 
quasar's spectra: the Ly-$\alpha$ forest. Remember that 
Lyman-alpha is the spectral line at 1216 \AA, in the far ultraviolet, 
that corresponds to the transition of an electron between the two 
lowest energy levels of a hydrogen 
atom (with the rest of the transitions giving rise to the 
whole Lyman series). 
 
Distant quasars get absorbed by 
many more clouds than nearby quasars. Quasars also emit a strong Lyman-alpha 
emission line. But the absorbing clouds all have 
smaller redshifts than the quasar since they have smaller distances. As a 
result the absorption lines are all on the blue or shorter 
wavelength side of the quasar emission line.\cite{elw1} 
 
\subsubsection{The Tully-Fisher relation} 
 
This is a promising distance determination technique (widely used in the 
70's and 80's). Once improved, it is now one of the more accurate secondary 
distance indicators for the Universe. It relies on the relationship between 
the rate at which a spiral galaxy spins and its intrinsic luminosity: 
the faster a galaxy spins, the more luminous it is. 
 Motion will cause a narrow line, e.g., a line due to some element like 
hydrogen, to be smeared out and to appear broad to the external 
observer. 
 
The broader the line, the faster the galaxy is spinning. 
 Centrifugal force and gravitational force are in balance 
\begin{equation} 
    \frac{  v^2}{ R} - \frac{G M_{galaxy}}{ R^2} = 0. 
\end{equation} 
And this leads to a spin 
rate  of the type: 
\begin{equation} 
    M_{galaxy} = \frac{v^2  R}{ G }. 
\end{equation} 
 
\subsubsection{The Hertzsprung-Russell diagram} 
 
The HR diagram is a plot of all known stars, where on the horizontal axis 
one puts 
the star surface temperature (decreasing to the right), and on the 
vertical one the star luminosity. 
In the early 1900's, Ejnar Herstzprung and Henry Norris Russell independently 
made the discovery that the luminosity of a star is related to its surface 
temperature. When luminosity versus temperature plots are made, stars do not 
fall randomly on the graph; rather they are confined to specific regions. 
 
This tells you that there is some physical relationship between the 
luminosity and temperature of a star. Herstzprung and Russell 
 made groupings of stars in the 
diagrams, according to the names: Main Sequence, Giants, Super-Giants, 
 and White Dwarfs (these groups are referred to as {\it luminosity classes}). 
Stars spend most of their lives as main sequence stars. 
 
The HR Diagram may be partially understood in terms of the luminosity for an 
 object emitting thermal radiation: 
\begin{equation} 
L \sim R^2T^4. 
\end{equation} 
If all objects in the HR diagram were the same size then all objects would 
lie along a diagonal line of slope = 2 in this logarithmic plot. 
Schematically, stars fall into regions above and below with respect to 
this line, which is actually deformed into a curved one 
containing the main sequence.

\subsubsection{The Hubble law} 
\begin{figure}[htb] 
\centerline{\psfig{file=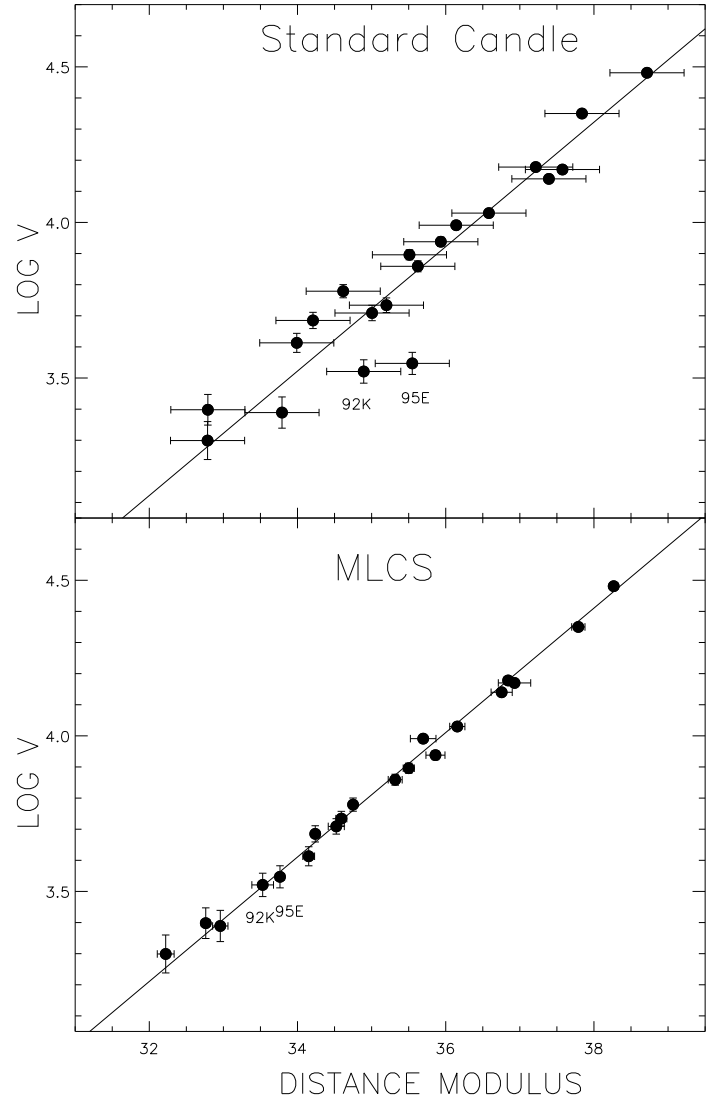,width=10cm}} 
\vspace*{20pt} 
\caption{A modern determination of Hubble's law, by a empirical method 
that uses multicolor light curve shapes 
(MLCS) to estimate the luminosity, distance, and 
    total line-of-sight extinction of Type Ia supernovae (SN Ia). 
 A. Riess, W. Press, and R. Kirshner, Astrophys. J. {\bf 473}, 88 (1996).} \label{twohub} 
\end{figure}

States that radial velocities of galaxies are proportional to their distance 
(a Doppler shift caused by the cosmic expansion, Fig. \ref{twohub}). 
Namely: 
\begin{equation} 
v = \frac{dD}{dt} = H D,  \qquad  v = cz + \cdots, 
\end{equation} 
being the redshift 
\begin{equation} 
1 + z =\frac{\lambda_{observed}}{\lambda_{emitted}} 
 = \sqrt{\frac{1+v/c}{1-v/c}}. 
\end{equation} 
 
The value for the Hubble constant at present, obtained from observations 
by WMAP is 
\begin{equation} 
H_0= 71 \pm 3.5 \ \mbox{km/sec/Mpc}. 
\end{equation}

One should note that the Hubble law stated as $v = HD$, that is, as 
 true for all values of $D$, even 
very large ones ($v > c$), must be modified in a curved spacetime, 
by establishing a chain of patches in each of which it is indeed 
applicable. In the way: 
\begin{equation} 
 D_{now} = D_{\mbox{\small us to Z}}= D_{\mbox{\small us to A}} 
+D_{\mbox{\small A to 
B}}+\cdots +D_{\mbox{\small X to Y}}+D_{\mbox{\small Y to Z}}. 
\end{equation} 
 The relation between the  Hubble law distance $D_{now}$, the 
velocity $ v$ and the  redshift $z$ is: 
\begin{equation} 
 v = H_0 D_{now}, \qquad  D_{now} = \frac{c}{H_0} \ln (1+z), \qquad 
 1+z = \exp (v/c). 
\end{equation} 
 While the Hubble law distance is (in principle) 
measurable, the need for helpers all along the chain of galaxies 
out to a distant galaxy (as described above) makes its use impractical. 
 
\subsubsection{Measurements of the distance} 
 
Other distances are measured more easily. One of them is 
 
\begin{itemlist} 
 \item {\it The angular size distance}: 
\begin{equation} 
     \theta = \frac{\mbox{size}}{D_A}, 
\quad \mbox{that is} \quad    D_A =  \frac{\mbox{size}}{\theta}, 
\end{equation} 
being here the size equal to the transverse extent of the object, and $\theta$ 
the angle (in radians) that it subtends on the sky. 
 
\item {\it The luminosity distance}, $ D_L$ is defined  through: 
\begin{equation} 
         \mbox{ Flux} = \frac{\mbox{luminosity}}{4 \pi D_L^2}. 
\end{equation} 
And the {\it light travel time}, by: 
\begin{equation} 
D_{ltt} = c (t_0-t_{em}). 
\end{equation} 
 
\item {\it Use of the Tully-Fisher relation.} 
This relation has been discussed before. The rotational velocity 
of a spiral galaxy is an indicator of its luminosity: 
\begin{equation} 
L = \mbox{Const} \  V_{rot}^4. 
\end{equation} 
The rotational velocity is measured using an optical spectrograph 
or radio telescopes and thus gives us the luminosity. Combined 
then with the measured flux, this luminosity gives the distance. 
 
Let us consider two galaxies: a giant spiral and a dwarf spiral. 
Say the small galaxy is closer to the Earth, so that both cover 
the same angle on the sky (i.e., both have the same apparent 
brightness). But we observe that the distant galaxy has greater 
rotational velocity, since the difference between the redshifted 
and the blueshifted sides is larger. Using this information, the 
relative distance of the two galaxies can be determined. 
 
\item {\it The Faber-Jackson relation.} 
The stellar velocity dispersion, $s(v)$, of the stars in an 
elliptical galaxy is an indicator of its luminosity. In fact: 
\begin{equation} 
L = \mbox{Const}\  s(v)^4. 
\end{equation} 
The velocity dispersion is measured using an optical spectrograph, 
and this gives us the luminosity. Combined with the measured flux, 
this luminosity gives the distance. 
 
\item {\it  Gravitational lens time delay.} This gives also a 
measure of the relative distance. A quasar viewed through a 
gravitational lens turns into  multiple images. But these light 
paths from the quasar to us have different lengths, that differ by 
\begin{equation} 
D [\cos(q_1)-\cos(q_2)], 
\end{equation} 
where $q$ is the deflection angle, and $D$ the distance to the 
quasar. 
 
And, since quasars are time variable sources, we can measure the 
path length difference by looking for time-shifted correlated 
variability in multiple images.\cite{lensing1} \end{itemlist} 
 
\subsubsection{Age of the Universe} 
\begin{itemlist} 
\item {\it  Age of the chemical elements.} Rubidium and strontium 
are usually found in rocks. But Rb$^{87}$ decays into Sr$^{87}$ 
(radiogenic) with a half-life of 47 billion years. Sr$^{86}$ is 
not produced by any rubidium decay (non-radiogenic), so that it 
can be used to determine what fraction of Sr$^{87}$ was produced 
by radioactive decay, by plotting the Sr$^{87}$/Sr$^{86}$ ratio 
versus the Rb$^{87}$/Sr$^{86}$ ratio.\cite{elw1} 
\item {\it Earth.} The oldest rocks are about 3.8 billion years old. 
The oldest meteorites are dated to be 4.56 billion years old. 
\item {\it Radioactive dating of old stars.} From the Thorium 
abundance in an old halo star: the Th/Eu (Europium) ratio in such 
star is 0.219 compared to 0.369 in the Solar System now. Th decays 
with half-life of 14.05 Gyr.\cite{cowan1} 
 This gives $15.6 \pm 4.6$ Gyr for the age, based on two 
stars (CS 22892-052 and HD 115444). 
\item {\it Age of the older star clusters.} When stars are burning hydrogen to 
helium in their cores, they fall on a single curve in the 
luminosity-temperature plot (the HR diagram, already discussed). 
From analysis of this behavior an age for our Universe greater than 12.07 Gyr, 
 with $95\%$ confidence, has been found. 
\item {\it Ages of white dwarfs.} In the globular cluster M4, it has 
been found to be $12.7 \pm 0.7$ Gyr. In 2004 Hansen et al.\cite{hansen1} 
 updated 
their analysis to find an age for M4 of $12.1 \pm 0.9$ Gyr, which 
is very consistent with the age of globular clusters from the main 
sequence. 
\item {\it Age of the Universe.} The current best value is 
$13.3 \pm 0.2$ billion years, from WMAP, but has been steadily 
increasing lately. Note however that the comoving radius of the 
universe is about 40 billion light-years (about a factor of 3 bigger). 
\end{itemlist} 
 
\subsubsection{The Universe is homogeneous and isotropic } 
 
The reader should not confuse homogeneity and isotropy. The 
pattern of a red brick wall (like the beautiful ones in Boston's 
Beacon Hill) is an homogeneous but not isotropic one. On the 
contrary, the pattern of light rays emitted in any direction by a 
shining light in darkness is isotropic but not homogeneous. 
 
Direct evidence for statistical homogeneity in the distribution of 
matter at sufficiently large scales 
came from the first accurate measurement of the galaxy 
two-point correlation function.\cite{sasl1} Totsuji and Kihara 
(1969)\cite{tk1} solved this long-standing problem. In their own words: 
{\it ``The correlation function for the spatial distribution of galaxies in 
the universe is determined to be $(r_0 / r)^{1.8}$, $r$ being the 
distance between galaxies. The characteristic length $r_0$ is 4.7 Mpc. 
This determination is based on the distribution of galaxies brighter 
than the apparent magnitude 19 counted by Shane and Wirtanen (1967). 
The reason why the correlation function has the form of the inverse 
power of r is that the universe is in a state of `neutral' stability."} 
Deep physical insight into the gravitational many-body problem ---usually 
a good way to short-circuit complicated mathematical formalism--- 
led Totsuji and Kihara to their conclusion. Previous guesses at 
an exponential or Gaussian form for the correlation had been intensively 
discussed. With the new results, galaxy clustering could be considered 
to be a phase transition from a Poisson distribution to a correlated 
distribution, slowly developing on larger and larger scales as the 
universe expands. 
 
\begin{figure}[!t] 
\centerline{\psfig{file=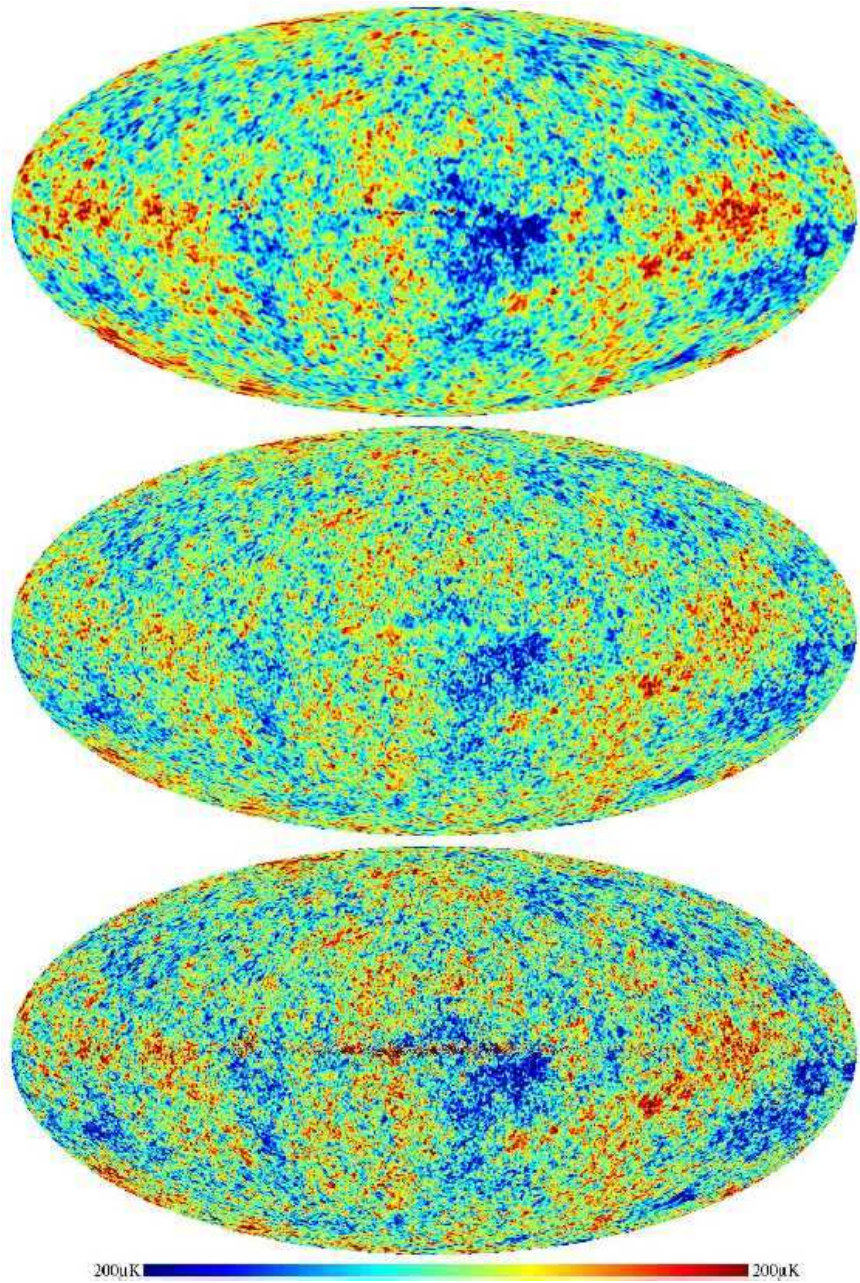,width=11.5cm}} 
\vspace*{3pt} 
\caption{The linearly cleaned WMAP map (top), 
a Wiener filtered map (middle) and the raw map (bottom). 
All maps are shown in Mollweide projection 
in galactic coordinates with the galactic center $(l,b)=(0,0)$ in the 
middle and 
galactic longitude $l$ increasing to the left. 
M. Tegmark, A. de Oliveira-Costa, and A. Hamilton, Phys. Rev. 
{\bf D68}, 123523 (2003).}  \label{cmb1} 
\end{figure}

The isotropic and homogeneous Universe case became much stronger 
after Penzias and Wilson announced the discovery of the Cosmic 
Microwave Background in 1965. 
In fact, as we now know, the deviations from 
homogeneity in the CMB radiation are of about a part in $10^5$, 
what makes it {\it too} homogeneous and creates a severe problem 
when one wants to find some inhomogeneities, to serve as  seeds 
for star and galaxy formation. Beautiful maps of the whole universe 
showing the temperature fluctuations of the CMB have been produced 
by WMAP (Fig. \ref{cmb1}). 
 
It is not easy to draw these maps. A lot of effects must be accounted 
for and conveniently substracted from the image: the dipole 
contribution, the galaxy plane contribution, observational biases, 
etc.). The figure shows three different stages of the production of 
a final map (the cleaned one, top of Fig.  \ref{cmb1}). 
 
\begin{figure}[!ht] 
\vspace*{-120pt} 
\centerline{\psfig{file=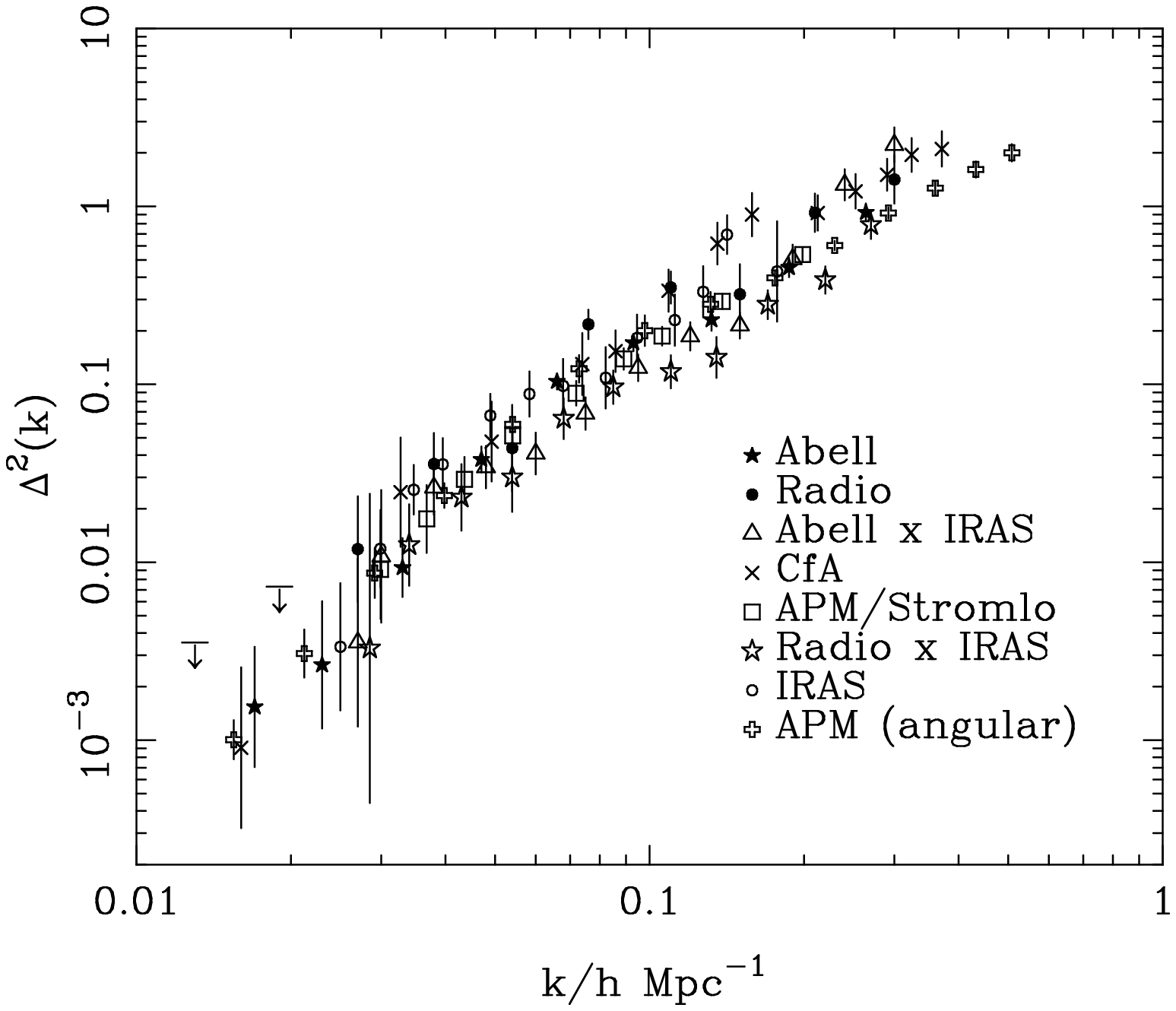,width=15cm}} 
\vspace*{-140pt} 
\caption{For 100 Mpc regions the Universe is smooth to within several 
percent. From J.A. Peacock and S.J. Dodds, Mon. Not. Roy. Astron. 
Soc., 267 (1994) 1020.}  \label{page2} 
\end{figure}

The following plot (Fig. \ref{page2}) shows that our universe 
approaches homogeneity (as measured now from the matter distribution) 
as big enough regions of the same are considered, of about or larger 
than 100 Mpc. 

\subsection{Short summary of inflation} 
 
Basic to cosmological observations, as those that already lead to 
the Big Bang model, is the consideration of a scale factor, 
$a(t)$, to be taken e.g. as the distance between any pair of 
comoving objects (e.g. two distant galaxies), or even the 
curvature of the universe itself, if it is non-vanishing. The 
scale factor grows by an amount $1+H dt$ during a time interval 
$dt$, that is: 
\begin{equation} 
 D_G(t) = a(t) D_G(to), 
\end{equation} 
with $D_G(to)$ being the  distance to the galaxy $G$ right now, 
and $a(t)$ a  universal scale factor that applies to all comoving 
objects. This law had to be changed for the description of the 
very beginning of the cosmos, owing to the serious problems of the 
original Big Bang model. 
 
A. Starobinsky and A. Guth  offered a solution to the 
flatness-oldness problem and to the horizon (or causality) problem of the old 
Big Bang theory, that was absolutely unable to explain them (together with 
some other, as the present absence of magnetic monopoles). 
In 1980,  Alan Guth 
proposed a modification to the Big Bang theory, by suggesting that in the 
first moments of its life our universe inflated as if it had been the soappy 
membrane of a small bubble, that become gigantic in a small 
fraction of a second.\cite{stef1} 
Inflation is in fact a modification of the conventional Big Bang 
theory, proposing that the expansion of the universe was propelled 
by a repulsive gravitational force generated by an exotic form of 
matter. Although Guth's initial proposal was flawed, this was soon 
overcome by the invention of ``new inflation," by Andrei Linde and 
independently by Andreas Albrecht and Paul Steinhardt. {\it ``The bang 
was there, but it was not big,"}  said at some occasion A. 
Linde.\cite{inflat1} 
 
Nowadays no self-respecting theory of the Universe is complete without a 
reference to inflation. But there is in the meantime 
 such a large variety of versions 
that it would be impossible here to provide a minimally consistent 
account even of the basic ideas to encompass all them, thus the 
reader is referred to the excellent  bibliography by the creators 
of the theory.\cite{inflat2} 
 
Inflationary theory does not replace Big Bang theory, but adds an extra stage: 
 before the Big Bang, the universe went through a period of extremely rapid 
 expansion, growing by 30 orders of magnitude in a fraction of a second. It 
is difficult to imagine something becoming this large this 
quickly. In the words of Guth: {\it ``To picture a pea expanding 
to the size of the Milky Way more quickly than the blink of an 
eye.''} When he came up with the theory of cosmic inflation, Guth 
was a 34-year old physicist at the Stanford Linear Accelerator 
Center, in the ninth year of a seemingly interminable career as a 
postdoctoral fellow. He was working on the problem of magnetic 
monopoles: the Big Bang model predicts an abundance of magnetic 
monopoles but none have ever been found. 
 
A few years earlier, Linde had suggested that, in its early stages, the 
universe had undergone a series of phase transitions, accompanied by 
supercooling. Supercooling is seen quite often in phase transitions from one 
form of matter to another, such as water cooling to ice. In supercooling, 
water will remain liquid as it cools below 0, but at the slightest 
disturbance it will immediately freeze. Guth and Tye were working on the 
problem of how supercooling in the early universe would affect the 
production of magnetic monopoles. 
{\it ``So I went home one night and did that calculation and discovered that it 
would have a dramatic effect on the evolution of the universe,"} Guth said 
once.\cite{stef1} 
The supercooled matter would cause gravity to reverse direction, so that 
objects would repel each other, resulting in exponential inflation. This 
would also make magnetic monopoles exceedingly rare. 
The impact of the theory was immediate. 
 
One major puzzle solved by inflation is the fact that it explains 
the extreme homogeneity and isotropy of the universe, as observed 
by COBE and now with much greater precision by WMAP and several 
balloons. This is a highly improbable state viewed from Big Bang 
theory. In the inflationary scenario, however, stretching out a 
tiny, uniform universe exponentially  results in a similarly 
uniform larger universe. Inflation also explains why parallel 
lines don't cross ---something everyone learns in school as a 
basic principle of Euclidean geometry. But other types of geometry 
are possible. The density of the universe determines whether it is 
open  or closed. Theoretical calculations show that a universe 
coming from the usual Big Bang should be very curved, whereas 
scientific observations show the universe as flat and Euclidean. 
This ``flatness" problem is also solved by inflation. For some 
interesting reference books see e.g. Ref. \cite{inflat3} 
 
One of the intriguing consequences of inflation is that quantum fluctuations 
in the early universe can be stretched to astronomical proportions, providing 
the seeds for the large scale structure of the universe. The predicted 
spectrum of these fluctuations was calculated in 1982. 
One thinks of vacuum as empty and massless (with a density 
$< 10^{-30}$ g/cc). Now, as we know from  Quantum Field Theory (QFT), the 
 vacuum is not empty but filled with virtual particles. 
These quantum fluctuations, once enormously enlarged by inflation, 
 can be seen today as ripples in the cosmic background 
radiation, but the amplitude of these faint ripples is only about 
one part in $10^5$. Nonetheless, these ripples were detected by 
the COBE satellite in 1992, and they have now been measured to 
much higher precision by the WMAP satellite and several balloons (like 
MAXIMA, DASI and BOOMERanG). The properties of the radiation are 
found to be in excellent agreement with the predictions of the 
simplest models of inflation. Also, according to Guth and 
Farhi,\cite{gfa1} with quantum tunneling it might be theoretically 
possible to ignite inflation in a hypothetical laboratory, thereby 
creating a new universe. The new universe, if it can be created, 
would not endanger our own universe. Instead it would slip through 
a wormhole and rapidly disconnect completely. And yet another 
intriguing feature of inflation is that almost all versions of 
inflation are eternal: once inflation starts, it never stops 
completely. Inflation has ended in our part of the universe, but 
very far away one expects that inflation is continuing, and will 
continue forever. Is it possible, then, that inflation is also 
eternal into the past? Recently Guth,  Vilenkin  and 
Borde\cite{gvb1} have shown that the inflating region of spacetime 
must have a past boundary, and that some new physics, perhaps a 
quantum theory of creation, would be needed to understand it. 
 
The increasing precision of cosmological data sets is opening up 
new opportunities to test predictions from cosmic inflation. The 
impact of high precision constraints on the primordial power 
spectrum is expected to be important and the new generation of 
observations could 
 provide real tests of the slow-roll inflation paradigm, as well as produce 
significant discriminating power among different slow-roll models. 
In particular, proposed next-generation measurements of the CMB 
temperature anisotropies, and specially polarization, as well as 
new Lyman-$\alpha$ measurements  could become practical in the 
near future. Relationships between the slope of the power spectrum 
and its first derivative  are nearly universal among existing 
slow-roll inflationary models, and therefore these relationships 
can be tested on several scales with new observations. Among other 
things, this provides additional motivation for the measure of CMB 
polarization, to be accomplished with the PLANCK mission, 
in which our group is participating. 
 
\subsection{On the topology and curvature of space} 
 
The Friedmann-Robertson-Walker (FRW) model, which can be derived 
as the only family of solutions to the Einstein's equations 
compatible with the assumptions of homogeneity and isotropy of 
space, is the generally accepted model of the cosmos (more details 
later in these lectures). But, as we surely know, the FRW is a 
family with a free parameter, $k$, the curvature, that can be 
either positive, negative or zero (the flat or Euclidean case). 
This curvature, or equivalently the curvature radius, $R$, is not 
fixed by the theory and should be matched with cosmological 
observations. Moreover, the FRW model, and Einstein's equations 
themselves, can only provide local properties, not global ones, so 
they cannot tell about the overall topology of our world: is it 
closed or open? finite or infinite? Even being quite clear that it 
is, in any case, extremely large ---and possibly the human species 
will never reach more than an infinitesimally tiny part of it--- 
the question is very appealing to any of us. Note that all this 
discussion concerns only three dimensional space curvature and 
topology, time will not be involved. 
 
\subsubsection{On the curvature} 
 
Serious attempts to measure the possible curvature of the space we 
live in go back to Gauss, who measured the sum of the three angles 
of a big triangle with vertices on the picks of three far away 
mountains (Brocken, Inselberg, and Hohenhagen). He was looking for 
evidence that the geometry of space is non-Euclidean. The idea was 
brilliant, but condemned to failure: one needs a much bigger 
triangle to try to find the possible non-zero curvature of space. 
Now cosmologist have recently measured the curvature radius $R$ by 
using the largest triangle available, namely one with us at one 
vertex and with the other two on the hot opaque surface of the 
ionized hydrogen that delimits our visible universe and emits the 
CMB radiation (some 3 to 4 $\times 10^5$ years after the Big 
Bang).\cite{mteg1} The CMB maps exhibit hot and cold spots 
(see Fig. \ref{cmb1}). It can 
be shown that the characteristic spot angular  size corresponds to 
the  first peak of the temperature power spectrum, which is 
reached for an angular size of $.5^o$ (approximately the one 
subtended by the Moon) if space is flat. If it has a positive 
curvature, spots should be larger (with a corresponding 
displacement of the position of the peak), and correspondingly 
smaller for negative curvature. 
 
The joint analysis of the considerable amount of data obtained 
during the last years by the balloon experiments (BOOMERanG, 
MAXIMA, DASI), combined also with galaxy clustering data, have 
produced a lower bound for $|R|> 20 h^{-1}$Gpc, that is,  twice as 
large as the radius of the observable universe, of about $R_U 
\simeq 9h^{-1}$Gpc.

\subsubsection{On the topology} 
 
Let us repeat that GR does not prescribe the topology of the 
universe, or its being finite or not, and the universe could 
perfectly be flat and finite. The simplest non-trivial model from 
the theoretical viewpoint is the toroidal topology (that of a tyre 
or a donut, but in one dimension more). This topology has been 
studied in depth by the author (more about that will come in the 
last lecture). Traces for the toroidal topology and more 
elaborated ones, as negatively curved but compact spaces, have 
been profusely investigated, and some circles in the sky with near 
identical temperature patterns were identified.\cite{ncorn1} And 
yet more papers appear from time to time proposing a new 
topology.\cite{lumi1} 
 However, to summarize all these efforts and the observational 
 situation, and once the numerical data are 
interpreted without bias (what sometimes was not the case, and led 
to erroneous conclusions), it seems at present that available data 
point towards a very large (we may call it {\it infinite}) flat 
space.

\subsection{Expansion around a given probability density function} 
 
This section will be rather more technical. 
It is the starting point of some 
original work that we produced aimed at the study of matter and 
temperature fluctuations. 
 
 For any two probability density functions (pdfs), $f^{(1)}(x)$ 
and $f^{(2)}(x)$, with cumulants $\kappa_J^{(1)}$ and 
$\kappa_J^{(2)}$, it turns out that we can always write one of the 
pdfs in terms of the other one. In fact, the explicit formula 
is:\cite{kso1} 
\begin{equation} 
f^{(1)}(x) = \exp\left[ \sum_{J=0}^\infty ~(-1)^J~{\kappa_J^{(1)}- 
\kappa_J^{(2)}\over{J!}} ~ {d^J\over{dx^J}}  \right] ~ f^{(2)}(x). 
\label{e51} 
\end{equation} 
Remember the definition of the cumulants, from the generating function 
\begin{equation} 
\ln \Phi(t) = \sum_{J=0}^\infty \kappa_J \frac{(it)^J}{J!}; 
\end{equation} 
it turns our that the cumulants (or connected moments) 
 are immediately related with the usual 
central moments of the distribution: 
\begin{equation} 
\kappa_1 =m, \quad \kappa_2 =\sigma^2, \quad \kappa_3 =\mu_3, \quad \kappa_4 
=\mu_4 - 3\mu_2^2, \ldots 
\end{equation} 
The proof of Eq. (\ref{e51}) is simple and I'll just give it as a 
chain of equalities that the reader may check with some care. \\ 
{\it Proof}: 
\begin{eqnarray} 
&& \hspace*{-17mm}\Phi^{(1)}(t) = < e^{itx}> = \int_{-\infty}^\infty e^{itx} f^{(1)}(x) \ dx 
= \int_{-\infty}^\infty e^{itx} \ \exp [ \ ] \ f^{(2)}(x) \ dx \nonumber \\ 
&& \hspace*{-6mm}  = \mbox{  [integration by parts]  } 
=  \int_{-\infty}^\infty  f^{(2)}(x) \ \exp [ \ ] \ e^{itx} \ dx \nonumber \\ 
&&  =  \int_{-\infty}^\infty  f^{(2)}(x)\ \exp \left[ \sum_{J=0}^\infty {\kappa_J^{(1)}- 
\kappa_J^{(2)}\over{J!}} (it)^J  \right]\ e^{itx} \ dx  \nonumber \\ 
&&  =  \exp \left[ \sum_{J=0}^\infty {\kappa_J^{(1)}- 
\kappa_J^{(2)}\over{J!}} (it)^J  \right]  \int_{-\infty}^\infty  f^{(2)}(x)\ 
e^{itx} \ dx \nonumber \\ 
&& =  \exp \left[ \sum_{J=0}^\infty {\kappa_J^{(1)}- 
\kappa_J^{(2)}+\kappa_J^{(2)}\over{J!}} (it)^J  \right]\nonumber \\ 
&& =  \exp \left[ \sum_{J=0}^\infty {\kappa_J^{(1)}\over{J!}} 
(it)^J \right]. \nonumber 
\end{eqnarray}

\subsection{Fluctuations of the density field} 
 
They have been described before: 
\begin{equation} 
\delta = \frac{\rho -\bar{\rho}}{\bar{\rho}} = 
\frac{\rho}{\bar{\rho}} -1, 
\end{equation} 
where the values are taken smoothed over some fixed scale, $R$. 
Here $\rho$ is the value of the density field, and we have that 
\begin{equation} 
0 \leq \rho < +\infty,\qquad   -1 \leq \delta < +\infty. 
\end{equation} 
The fluctuation $\delta = \delta (\vec{r})$ is a stochastic 
field ($\vec{r}$ position) 
 while $p(\delta)$ is the one-dim probability 
density function  of this random variable. What is very 
important to note here is the fact that the density field 
fluctuation, $\delta$ is a random variable which is 
{\it bounded below} by the 
value $-1$ (and thus highly asymmetric), together with the fact 
that, on the other hand, the Gaussian pdf corresponds to a 
stochastic variable which is unbounded and symmetric. 
 
Now the main problem is here to: 
\begin{itemize} 
\item 
{\sl Recover the full shape of $p(\delta)$ from a few first order 
moments} \end{itemize} 
where we understand that it is the optimal solution to this 
question the one that is most interesting: using the minimal 
number of moments we want to obtain the best possible 
representation of the pdf. 
 
 Its solution is in fact {\it not 
unique}. Several approaches haven been historically considered: 
\begin{itemize} 
\item The {\it Zeldovich approximation}, which 
\begin{itemize} 
\item derives $p(\delta)$ from analytic approximations; 
 \item reproduces important aspects of the non-linear 
dynamics, but it yields a \item poor approximation  for the pdf 
and the  moments. 
\end{itemize} 
\item The {\it improved Zeldovich approximation}, which 
\begin{itemize} 
\item is an exact non-linear perturbation theory for the moments; 
\item is used for deriving the pdf from the Edgeworth expansion; 
and 
\item has an accuracy proportional to the order of the cumulants involved. 
\end{itemize} 
\item The {\it Edgeworth expansion}, which is obtained in terms of Gaussians and 
has been used to deal with  matter pdfs and also with  CMB fluctuations. 
Other uses of the Edgeworth expansion can be found in  biology, 
 economy and finance, and  mathematical statistics. But it 
has some shortcomings, in practice, as the appearance of negative 
probability values, in particular it  assigns $\neq 0$ probability 
to $\delta < -1$, as a consequence of the fact that a Gaussian pdf 
only makes sense for $\sigma < 1$. \item Alternative expansions 
are: 
\begin{itemize} 
\item The {\it Gamma} (or {\it Pearson type-III}) pdf. 
\item Other distributions. 
\end{itemize} 
\end{itemize}

\subsection{Saddle point approximation and the Edgeworth expansion} 
 
Gravitational clustering from Gaussian initial 
conditions predicts that 
\begin{equation} 
{ <\delta^J >_{c}} \propto { <\delta^2 >_{c}}^{J-1}, 
\end{equation} 
on large-scales (where the subindex $c$ indicates `connected' 
terms, which correspond to the cumulants), thus the following 
ratios are most interesting, and widely used in cosmological 
analysis: 
\begin{equation} 
S_{J}\, \equiv \,\frac{\kappa_J}{{\kappa_2^{J-1}}} = 
 { < \delta^J >_c \over { < \delta^2 >_c}^{J-1}}. 
\end{equation} 
They are called {\it hierarchical coefficients}, being $S_3$ the 
{\it Skewness} and $S_4$ the {\it Kurtosis}. 
 
To obtain an asymptotic expansion of $p(\delta)$ for small $\delta$, the 
Legendre transform is most appropriate: 
\begin{equation} 
\bar{\delta} \equiv d\psi(t)/dt, \quad  G(\bar{\delta}) = \bar{\delta} t - 
\psi(t), 
\end{equation} 
\begin{equation} 
p(\delta) =  \int_{G'=-i \infty}^{G'= +i \infty} \hspace{-1mm} 
{{G'' d \bar{\delta}} \over {2 \pi}} \exp[{-\delta 
G'(\bar{\delta})\,+\,\bar{\delta}G'(\bar{\delta})\,-\, 
G(\bar{\delta})}]. 
\end{equation} 
It is dominated by stationary points of the exponential at 
$\delta=\bar{\delta}$. The saddle point approximation yields: 
\begin{equation} 
p(\delta) = {{[G''(\delta)/{2\pi}]^{1/2}\, \exp[-G(\delta)]} \over 
{\int_{- \infty}^{ + \infty}{[G''(\delta)/{2\pi}]^{1/2}\, 
\exp[-G(\delta)]d \delta}}}. 
\end{equation} 
From the generating function, $\psi(t)$, 
\begin{equation} 
\psi(t)= \sum_{n=2}^{\infty} {{\mu_n} \over {n!}} t^{n} 
 = {1 \over 2} \sigma^2 t^2 + {1 \over 6}S_3 \sigma^4 t^3 + 
{1 \over 24}S_4 \sigma^6 t^4 + \cdots, 
\end{equation} 
one readily has 
 \begin{equation} 
G(\delta) \approx \left[{1 \over 2}\delta^2\,-\,{S_3 \over 6} \delta^3\,+\, 
{1\over 8} 
\left({S_3}^2 \,-\,{S_4 \over 3} \right)\delta^4 +\,{\cal O} (\delta^5) 
 \right]\, {\sigma^{-2}}, 
\end{equation} 
and finally 
\begin{eqnarray} 
p(\nu) &\simeq& \left\{ 1\,+\,{S_3 \over 3!}H_{3}(\nu)\sigma\,+\,\left[ 
{1 \over4!} 
S_4 H_4 (\nu) \right.\right.  \nonumber \\ 
&& \left.\left. + \,{10 \over 6!} {S_3}^2 H_6 (\nu) 
\right]{\sigma}^2 \right\} \, p_G (\nu) +\,{\cal O} 
\left({\sigma}^{3}\right), 
\end{eqnarray} 
the $H_n (\nu)$ being Hermite's polynomials: 
\begin{eqnarray} 
H_3(\nu) &=& \nu^3\,-\,3\nu, \nonumber \\ 
H_4(\nu) &=& \nu^4\,-\,6\nu^2\,+\,3, \nonumber \\ 
H_5(\nu) &=& \nu^5\,-\,10\nu^3\,+\,15\nu, \nonumber \\ 
H_6(\nu) &=& \nu^6\,-\,15\nu^4\,+\,45\nu^2\,-\,15, \nonumber \\ 
H_7(\nu) &=& \nu^7\,-\,21\nu^5\,+\,105\nu^3\,-\,105\nu, \nonumber \\ 
H_9(\nu) &=& \nu^9\,-\,36\nu^7\,+\,378\nu^5\,-\,1260\nu^3+945\nu, \nonumber \\ 
&\vdots& 
\end{eqnarray} 
This is the well-known (perturbative) Edgeworth series of a pdf up to 
third order. 
 Higher orders in the Edgeworth series can be obtained by keeping 
higher orders in the Taylor expansion.

\subsection{The Gamma ---negative binomial or  Pearson Type III 
(PT3)--- distribution} 
 
It arises from  the chi-square distribution with $N$ 
degrees of freedom when $1/\sigma^2=N/2$ is taken to be a continuous 
parameter 
\begin{equation} 
 \phi(\delta)  \equiv {(1+\delta)^{\sigma^{-2}-1} 
\over \sigma^{2\sigma^2} \Gamma(\sigma^{-2})} 
~\exp{\left(-{1+\delta\over{ \sigma^2}}\right)}. 
\end{equation} 
The hierarchical coefficients are constant: 
  $S_J=(J-1)!\,$  Note that the variable is in this case bounded 
from below, and that we can adjust the parameters in order that the 
bound be placed at $-1$. Using now the general result obtained 
before, we can easily get an expansion of any pdf in terms of the 
Gamma distribution, as follows\cite{gfe1} 
\begin{equation} 
p(\mu ) \equiv \sum_{n=0}^\infty c_n L_n^{(p-1)} (\mu ) \phi (\mu 
), 
\end{equation} 
with coefficients 
\begin{equation} 
c_n= \frac{n!\Gamma (p)}{\Gamma (n+p)} \int_0^\infty p(\mu ) 
 L_n^{(p-1)} (\mu )\, d\mu, 
\end{equation} 
being 
\begin{eqnarray} 
\phi(\mu)d\mu = {1 \over {\Gamma(p)}}\mu^{p-1} e^{-\mu}d\mu, 
\qquad \qquad  \mu = {x\,-\alpha \over \beta}\, \geq 0. 
\end{eqnarray} This is a three-parameter ($p,\alpha$ and $\beta$) 
family of distributions out of which only one, $p$, is relevant 
for normalized variables (such as the density fluctuations, 
$\delta$). Now everything is written in terms of the generalized 
Laguerre polynomials 
\begin{equation} 
L_n^{(p-1)} (\mu ) = \sum_{k=0}^n \frac{(-1)^k}{k!} 
\displaystyle{\left(_{\ n-k}^{n+p-1} \right)} \mu^k, 
\end{equation} 
which are, in particular, 
\begin{eqnarray} 
L_1^{(p-1)} (\mu ) &=& p-\mu, \nonumber \\ 
L_2^{(p-1)} (\mu ) &=& 
\frac{p(p+1)}{2}-(p+1)\mu +\frac{\mu^2}{2}, \nonumber \\ 
L_3^{(p-1)} (\mu ) &=& 
\frac{p(p+1)(p+2)}{6}-\frac{(p+1)(p+2)}{2}\mu \nonumber \\ 
&+&\frac{p+2}{2}\mu^2 -\frac{\mu^3}{6}. 
\end{eqnarray} 
The coefficients $c_n$ are found to be 
\begin{eqnarray} 
c_0 &=& 1, \quad c_1 =c_2 =0, \nonumber \\ 
c_3 &=& 
 -\frac{\Gamma (p+1)}{\Gamma (p+3)} (S_3 -2!), \nonumber \\ 
c_4 &=& \frac{\Gamma (p+1)}{\Gamma (p+4)} \left[ (S_4-3!)-12(S_3 -2!)\right], 
\nonumber \\ 
c_5 &=&  \frac{\Gamma (p+1)}{\Gamma (p+5)} 
\left[- (S_5-4!) + 20 (S_4 - 3!)-120(S_3 -2!)\right] ,  \nonumber \\ 
&\vdots & \nonumber \\ 
c_n&=& \frac{\Gamma (p+1)}{\Gamma (p+n)} \sum_{k=3}^n \left\{(-1)^k a_{n,k} 
 \left[ S_k -(k-1)!\right] \right. \nonumber \\ 
&& \left. + b_{n,k} 
 \left[ S_k -(k-1)!\right]^2+ \cdots \right\} 
\end{eqnarray} 
\begin{figure}[!htb] 
\centerline{\psfig{file=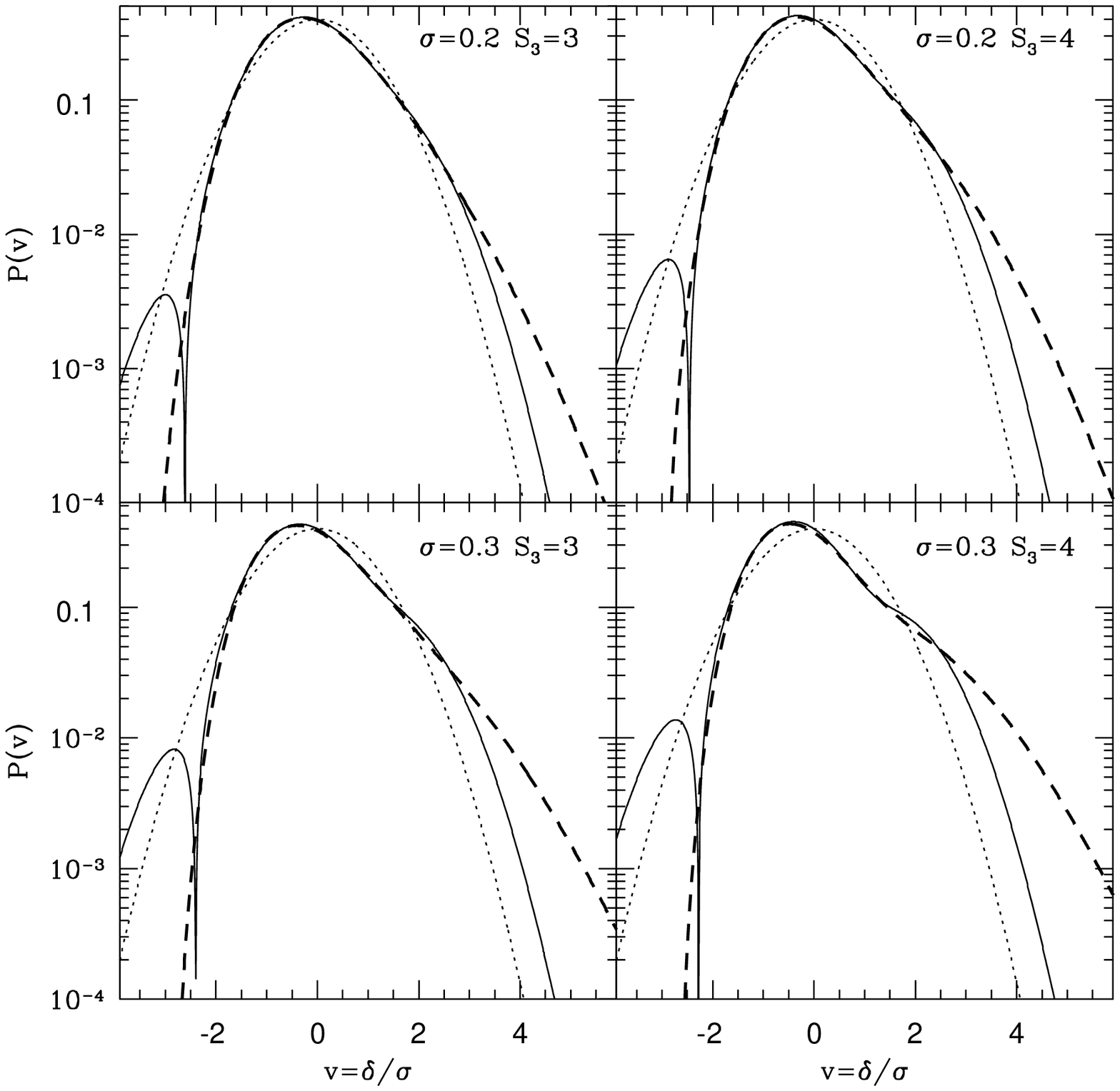,width=7.5cm} 
\psfig{file=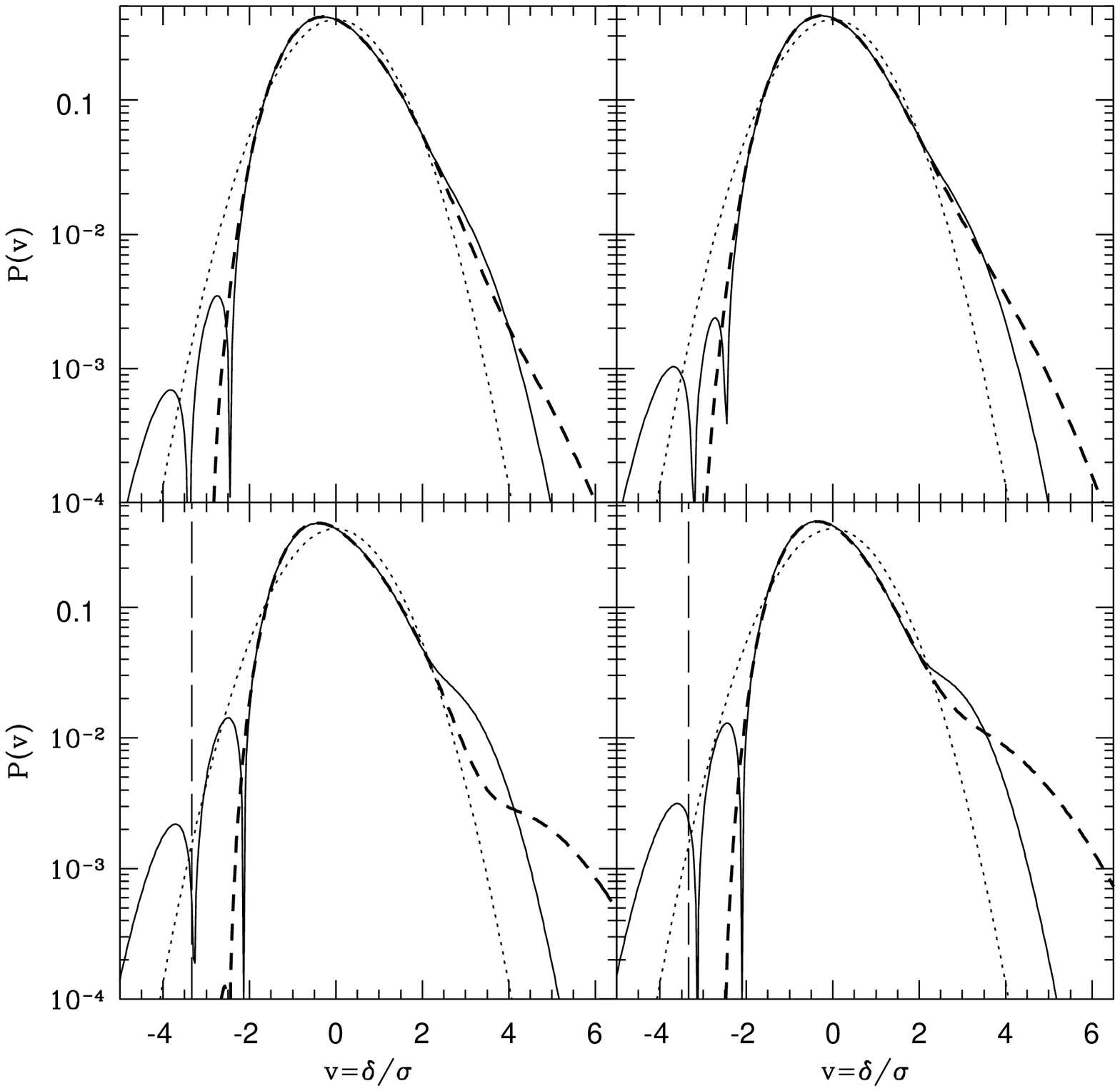,width=7.5cm}} 
\caption{Comparison of the leading order (left panel) and second 
order (right panel) Edgeworth  and  Gamma PDF expansions as 
functions of $\nu \equiv \delta / \sigma $ for several values of 
$\sigma$, and $S_J$ as labelled in the Figures. The  dotted, 
dashed and continuous lines correspond to the Gaussian PDF, Gamma 
and Edgeworth expansions, respectively. \label{efg1}} 
\end{figure} 
For a consistency check, it is clear that for $S_k=(k-1)!$ we 
recover the Gamma pdf: 
 $p(\mu) = \phi (\mu)$.    Finally 
\begin{eqnarray} 
p(\nu) &=&  \left\{ 1 + \sum_{n=3}^\infty \sigma^{n-2} \, F_n \, 
\right. \nonumber \\ && \times \left. \sum_{k=3}^n (-1)^{n-k} 
a_{n,k} [S_k-(k-1)!]\right\} \phi (\nu) \end{eqnarray} 
 
\begin{figure}[!htb] 
\centerline{\psfig{file=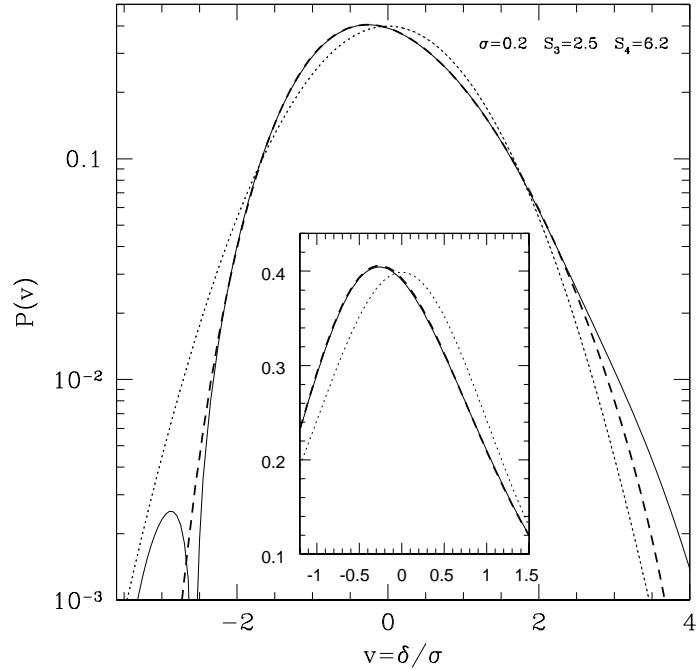,width=10cm}} 
\caption{Comparison of gravitational simulations with the second 
order Edgeworth and Gamma PDF expansions as a function of $\nu 
\equiv \delta/\sigma$. We use as parameters the measured values of 
$\sigma$, $S_3$ and $S_4$ as labeled in the Figure. The dotted, 
dashed and continuous lines correspond to the Gaussian, Gamma and 
Edgeworth distribution expansions. The inset shows a detail around 
the peak in linear scale. \label{efg2} } 
\end{figure}

Let us now compare the {\it Gamma} and the {\it Edgeworth} 
expansions: 
\begin{itemize} 
\item The Gamma expansion recovers all the terms 
that appear in the Edgeworth expansion plus some corrective terms. 
\item The Gamma expansion has 
exponential tails and a better general behaviour than the 
Edgeworth expansion, both on the positivity of $p(\nu)$ and the 
variate itself, $\rho$ (Fig. \ref{efg1}). 
\item The behavior at the pick is also improved, as a detailed 
analysis shows (Fig. \ref{efg2}). 
\end{itemize} 

\subsection{Non-Gaussian bounds from CMB observations} 
 
A basic ingredient in order to understand the  formation of large 
scale structures in our Universe is the  distribution of  initial 
conditions:  Have fluctuations been generated in the standard 
inflationary epoch? or  Do they require topological defects or 
more exotic assumptions? Generically, the first possibility leads 
to a Gaussian distribution, while the second leads to 
non-Gaussianities. 
 
This issue can be addressed through: 
\begin{itemize} 
\item Present day Universe fluctuations traced by the galaxy 
distribution. 
\item The presence of anisotropies in the 
cosmic microwave background (CMB). 
\end{itemize} 
An important contribution to the uncertainties comes from the sample 
variance (i.e., the finite size of the observational sample). In fact a 
 non-Gaussian signal can produce different 
sampling errors 
 
Now the problem is to: 
\begin{itemize} 
\item 
 {\it Place bounds on the degree of 
non-Gaussianity.} 
\end{itemize} 
 And the strategy to solve it consists in 
finding as many  independent results as possible in order to have 
a large sampling over the underlying distribution, that is 
specifically, to study the sample variance of CMB experiments over 
independent sky regions or subsamples, and also to perform a 
 Chi-square analysis taking different number of points and criteria. 
 
\begin{figure}[!thb] 
\centerline{\psfig{file=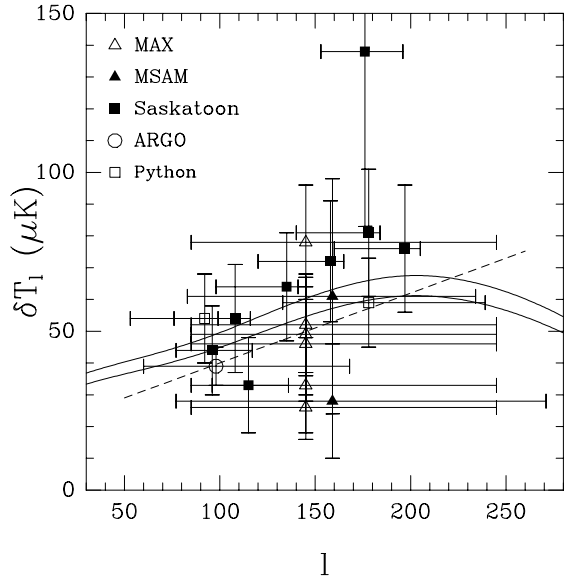,width=8cm}} \vspace*{1pt} 
\caption{Band power estimates of the {\it rms} temperature 
anisotropy $\delta T_l$ for observations given in our analysis. 
The vertical error bars show the (symmetrized) total errors in 
$\delta T_l$ while the horizontal ones stand for the width of the 
windows. The dashed line is the best fit slope to the data, 
$\delta T_l = (11/50) l_e + 18$. Continuous lines show the 
standard CDM model for two normalizations: $Q_{rms} = 20\, \mu$K 
(top), and $18\, \mu$K (bottom).} \label{cmbchi} 
 \end{figure} 
 
In our pioneering analysis,\cite{gfe2} performed some time ago, we 
started with a very few tests, namely Saskatoon, MAX, Phyton, 
MSAM, and ARGO, while subsequent work by our group and others has 
used lots of new data. The data points with their errors 
(horizontal ones corresponding to the window width) are displayed 
in Fig. \ref{cmbchi}.

What we found in our first study (which attracted considerable 
attention and was discussed in an editorial section of {\it 
Science}) was that, with the data at hand then, the Gaussianity 
hypothesis could be rejected at $\sim 80\%$ confidence level. We 
where careful, however, to point out at the fact that we had still 
very poor statistics and that maybe systematic errors in the 
experiments could had been underestimated, among other 
considerations\cite{gfe2}. In any case, we noted that by doubling 
systematic errors  we still obtained $\sim  60\%$ confidence. 
Subsequent analysis have diminished this predictions, 
 although the case for possible non-Gaussianities is 
far from being closed.\cite{ngwmap1} 
 Still more date will be needed to settle 
this issue. For interesting references on the subject of 
statistical analysis of the large scale galaxy distribution, 
together with a review of redshift galaxy surveys and the most 
recent determination of the cosmological parameters from these 
analysis, see e.g.\cite{sgd1}.

\section{Vacuum energy and the cosmological constant}

\subsection{The cosmological constant} 
 
 Our universe seems to be 
spatially flat and  to possess a non-vanishing cosmological 
constant. Thus, Einstein's `great mistake' may turn out ultimately 
to be a great discovery, a necessary ingredient in order to 
explain the acceleration of the universe. In any case, for 
elementary particle physicists it constitutes (in the words of J. 
Bj{\o}rken) a great embarrassment,\cite{bjork1} calculations there 
being off (when compared with physical facts) by the famous 120 
orders of magnitude. 
 
First, physicists tried to find a way  to get rid of it (Coleman, 
Weinberg, Polchinski, ...),\cite{cwp1} in the hope that it could 
be proven to be zero, what was hard enough. But now it turns out 
that it is non-vanishing, albeit very small, indeed a very 
peculiar quantity. 
 
The cosmological constant has to do with cosmology, of course 
(through  Einstein's equations and the  FRW universe obtained from 
them),\cite{weinb1} but it has  to do also  with the local 
structure of elementary particle physics as the  stress-energy 
density $\mu$ of the vacuum 
\begin{equation} 
L_{cc} = \int d^4x\, \sqrt{-g}\, \mu^4 = \frac{1}{8\pi G} \int 
d^4x\, \sqrt{-g}\, \lambda. 
\end{equation} 
In other words: two contributions appear, on the same footing 
\begin{equation} 
\frac{\Lambda \ c^2}{8\pi G} \  + \ \frac{1}{\mbox{Vol}} \, 
\frac{\hbar \, c}{2} \sum_i \omega_i. 
\end{equation} 
 
\subsection{From General Relativity to Cosmology} 
 
\subsubsection{On the meaning of Einstein's equations} 
 
Recall Einstein's equations (formulated in 1915-17), including a 
cosmological constant $\Lambda$: 
\begin{equation}  G_{\mu\nu}   - \Lambda g_{\mu\nu} = 
-8\pi G T_{\mu\nu}.  \end{equation} with 
\begin{equation} G_{\mu\nu}  \equiv R_{\mu\nu} - 
\frac{1}{2}g_{\mu\nu}R, \quad R =R_\mu^\mu. 
 \end{equation} 
These equations can be obtained from a variational principle, 
starting from an effective Einstein-Hilbert action 
\begin{equation} 
S= \frac{1}{16\pi G} \int d^4 x \, \sqrt{-g}\,  (R-2 \Lambda ) + 
\int d^4 x \, \sqrt{-g}\, {\mathcal L}_{mat}. 
\end{equation} 
They have a very profound meaning. 
 
On April this year I had to explain some couple of aspects of 
Special and General Relativity to the mixed TV audience of a 
Thursday evening. I decided to start with the most famous equation 
$E= cm^2$. {\it``Look''}, I said to the imposing camera in front 
of my face, {\it ``with this equation Einstein's genius put on the 
same footing matter and energy, which had been always thought to 
be non-matching quantities.  All of us have been told at primary 
school that apples and oranges don't match: no way to add 3 apples 
and 4 oranges. Since, what the result would be? Seven, ... but 
seven what? However, we go to the grocer's or the supermarket 
every week and we buy not only fruit, but all sort of different 
things, and the owner or the cashier just puts all items inside a 
bag and then ... he does it! precisely what we were told it was 
impossible to do: adds for us everything together and says `this 
makes \$14.76'. But this is {\em exactly} what Einstein did: 
to find a {\em conversion factor} for the different quantities, 
which in his case was the velocity of light and for the cashier 
it's just the price per pound of every item. Easy, isn't it? 
Einstein, as the grocer's, was not constrained by what we learn at 
school. In Einstein's case this opened the way to the possibility 
of converting matter into energy, and vice-versa, what was soon 
put to test.} 
 
And then I went on, {\it ``Now let's turn to GR. This is actually 
rather more difficult to grasp 
 and no shop owner or cashier on Earth would 
have guessed the answer this time. Look, in the language of 
Cosmology, Einstein's equation reads: $\Omega_{matter} 
+\Omega_{radiation} + \Omega_{\Lambda} + \Omega_{curvature} =1.$ 
This equation is not so widely known as the previous one, 
but is in no way less important. At first sight, you would say 
---alas, that's again the same as before, only more items appear 
say, a 
computer, a car, a cellular phone, a futon, a T-shirt, ..., 
of course one can buy them all together at the mall, no problem--- 
But this wouldn't be the whole truth. An important issue is 
missing from that argument, namely the last term, 
$\Omega_{curvature}$, which refers to the {\em mathematical} 
curvature of space-time itself. That's very different from the 
rest of terms, since it means that the {\em reference system} 
itself gravitates, that there is {\em no} `outside reference 
system'. In other words and following with the same example as 
before, what Einstein did here was to put the grocer himself 
inside the bag together with the rest of the things we bought! Who 
now will do the sum for us? Actually the first to guess our 
Universe could behave in this remarkable way was Ernst Mach 
(1838-1916).\footnote{There is a nice essay from Frank 
Wilczek on that issue.\cite{Wilczek1}} 
And Einstein found out the precise 
equations to try to confirm such extraordinary idea. This is what 
Gravity Probe B is going to confirm, with very good precision, so 
that there can be no doubt that our Universe does in fact behave this 
way.''} Then I went on, to explain frame warping and frame dragging.
\medskip 
 
What Einstein did, specifically, when building his equations: 
\begin{itemize} 
\item Geometry (curvature), radiation energy, matter, the cc, all 
are on the same footing and can be equated together. This is the 
mathematical concretion of Mach's principle. \item $ G_{\mu\nu}$ 
is a linear combination of the metric \ $ g_{\mu\nu}$   and of 
first and second derivatives of the same. \item $T_{\mu\nu}$ is 
the energy-momentum tensor, and $\Lambda$ a (possible) 
cosmological constant. 
\end{itemize} 
Actually Einstein didn't quite succeed in pining down in his 
theory of GR the whole content of Mach's principle (see 
\cite{Wilczek1}); but there is no doubt that 
remarkable glimpses of it are to be found in Einstein's equations, 
namely frame warping and frame dragging by distributions of matter 
and rotating massive shells, respectively. 
 
\subsubsection{Gravity Probe B} 
 
The mission Gravity Probe B was launched by NASA on April 20, 
2004, with the idea to try to see these two effects in great 
precision.\cite{gpb1} \begin{romanlist} 
\item {\it Frame warping} was proposed by deSitter 
in 1916, as the geodetic force a gyroscope would suffer in the 
presence of the space-time curvature induced by the presence of a 
mass. In the case of Gravity Probe B, which describes a polar 
orbit of 640 Km radius ---the gyroscopes' axis  having been 
oriented towards a convenient guide star (IM Pegasi)--- the 
calculated effect will be a displacement of the gyroscopes'
 orbit (it won't be exactly circular around 
the Earth) of 6.6 arcsec/year, with 
a expected  error of less than $10^{-4}$. 
 
\item {\it Frame dragging} was discovered by Lense and Thirring in 1918 
as a gravito-magnetic force.\cite{lensti1} It will be produced, 
 in the case of 
Gravity Probe B, by the Earth's (a massive body) rotation on the 
reference system, defined in this situation by the gyroscopes 
themselves. The orbit has been specifically chosen so that both 
effects on the gyroscopes are perpendicular. Frame dragging will 
result in the rotation axis of the gyroscopes trying to approach 
the Earth's rotation axis by an amount of 42 milliarcsec/year, 
with an estimated error of $10^{-2}$ (equivalent to the section of 
a human hair seen from 15 Km distance, while the effect amounts to 
seeing the same hair from 400 m. Reportedly, this precision is 
still unprecedented in experimental observations. More details 
about this important mission can be found in its web 
page,\cite{gpb1} where one can learn a lot about the physical 
meaning of GR, and this in the best possible way, namely, in the 
framework of an actual experiment. 
\end{romanlist} 
Although the idea is very simple, 
and the first plans to launch such a satellite were met over 40 
years ago, the technical difficulties involved are extraordinary 
and have postponed its final launch till a few months ago. 
 
If the results confirm what almost every physicist believes, i.e. 
 the validity and accuracy of GR, then there will be no way 
out but to admit that 
the mere notion of the existence of an  ideal reference frame in 
the cosmos is absolutely erroneous. Any mathematical reference 
will also `gravitate', that is, it will be unavoidably subject to 
the influence of all the masses in our Universe, and their 
rotation. In plain words, {\it `the grocer himself will have to be put 
into the bag, indeed, and nobody will be able to do the sum for 
us.'} This is what we learn from looking at our Universe. If 
confirmed, these results will completely demolish Isaac Newton's 
original formulation of the concept of {\it absolute space}, that was 
more clear to him than the purest of waters (and also to more one 
reputed philosopher, as I. Kant). 
 
\subsubsection{Solutions to Einstein's equations} 
 
The  Schwarzschild solution (1916)\cite{ksch1} of Einstein's 
equations reads 
\begin{equation} \hspace*{-4mm} ds^2 = \left( 1-\frac{2MG}{r} \right) dt^2 - 
\left( 1-\frac{2MG}{r} \right)^{-1} dr^2 -r^2d\theta^2 
-r^2\sin^2\theta d\varphi^2. \end{equation} It was soon realized 
that it could describe a black hole, but not a entire universe. 
However, the Friedmann-Lema\^{\i}tre-Robertson-Walker (1935-36) 
solution (FRW),\cite{frw1} first found by A. Friedmann in 1922, 
\begin{equation} ds^2 = dt^2 -R^2(t) \left( \frac{dr^2}{1-k r^2} +r^2d\theta^2+ 
r^2\sin^2\theta d\varphi^2 \right), \end{equation} with 
$k=0,+1,-1$, not only can do so, but it is the only solution of 
Einstein's equations (up to the constant $k$, the curvature) that 
can satisfy the requirements of homogeneity and isotropy, that our 
universe is (observationally) known to possess, to a high degree of 
accuracy. Now, in Cosmology the Friedmann equation is always 
written under the equivalent form: 
\begin{equation} 
{\dot a^2 \over a^2} = H^2_0 \left[ \Omega_R \left( {a_0 \over a 
}\right)^4 + \Omega_{NR} \left( {a_0 \over a } \right)^3 + 
\Omega_V + (1-\Omega)\left(\frac{a_0}{a}\right)^2\right], 
\end{equation} 
with  $\Omega_R$ being the radiation (or relativistic matter) 
content of our Universe, which satisfies the equation of state 
pressure = density/3, that is $ p_R = \frac{1}{3}\rho_R$, being 
$\rho_R \propto a^{-4}$, and $a$ a typical distance (say the mean 
intergalactic one), as the Universe expands; the ordinary 
(non-relativistic) matter is $\Omega_{NR}$, and satisfies $ p_{NR} 
= 0$, being $\rho_{NR} \propto a^{-3}$; the vacuum energy density, 
$\Omega_V$ is undistinguishable from the cosmological constant, 
with an equation of state $p_V = -\rho_V$, being here $\rho_V = 
{\rm const}$. Finally, the equation has been normalized to one, so 
that the sum of the different contributions equals this number; in 
other words, $1-\Omega$ in the equation before (where $\Omega= 
\Omega_R+ \Omega_{NR}+\Omega_V$) is the contribution of the 
geometry (of the curvature of the Universe, while $\Omega$ is 
`physical' contribution), $\Omega_K$, which behaves as $\rho_K 
\propto a^{-2}$, and 
\begin{equation}  \Omega_R + \Omega_{NR} + 
\Omega_K + \Omega_V =1.\end{equation} Presently, the contribution 
of $\Omega_K$ can be neglected, since from observations we obtain 
that $\Omega_K \simeq 0.0 \pm 0.1$, so that, as we explained 
before, we seem to live in a flat universe. Thus, one gets the 
famous {\it cosmological triangle}, which is a simple and 
intuitive graphical representation where one can read the 
proportions of the different terms in any proposed model for the 
present universe. The most plausible ones at present (e.g. from 
WMAP), yield a 
mere $4\%$ for the entire ordinary matter+energy content (i.e., 
barions+photons, with just some $.05\%$ for radiation), some $25\%$ 
comes from dark or invisible matter (trapped mainly in galaxy 
clusters, and a fraction in galaxy halos),\cite{darkm1} 
 and the biggest part, around $70\%$, is an absolutely unknown 
called dark energy (those values are for an $h \simeq .72$). 
 
Concerning dark matter, Zwicky noticed in 1933 already, that the 
gravitational action of the luminous matter was not enough in 
order to hold galaxy clusters together (could explain kpc 
structures at most.) Different kind of cold matter particles, with 
magnitudes that can differ in almost 100 orders, have been invoked by 
existing models, from  axions and neutrinos to large planets (or 
`Jupiters'), and also theories deviating from ordinary Newtonian 
physics. A portion of particle dark matter is sure to exist: the 
mass coming from neutrinos could be already 
 as large as the mass in visible 
stars. The lack of Newtonian matter is seen to occur at an extense 
range of distances, from the less than 1 kpc corresponding to 
dwarf spirals to the more than 100 Mpc in large clusters of galaxies. 
Twenty years ago, Milgrom made the remarkable observation that 
the need for dark matter in galaxies only arises when the Newtonian 
acceleration is less than a value $a_0\simeq .3 c H_0$. This is 
called now Milgrom's law, and has given rise to a theory, the Modified 
Newtonian Dynamics (MOND).\cite{darkm1} 
As we have seen before in some detail, 
of the known quantities only the cosmological 
constant has an equation of state which could contribute to the dark 
energy term (or, similarly, the proposed models of dynamic cc's, as 
quintessence\cite{quint1}). 
 
\subsubsection{In terms of the redshifts} 
 
An equivalent (and even more commonly used) expression is in terms 
of the redshift, $z$ (taken in practice as the inverse of the 
cosmological time) 
\begin{eqnarray} 
d_H(z) = H_0^{-1} \left[ \Omega_R (1+z)^4 + \Omega_{\rm NR} 
(1+z)^3 + (1-\Omega) (1+z)^2 + \Omega_V\right]^{-1/2}. 
\end{eqnarray} 
Limiting forms: 
\begin{eqnarray} 
d_H(z) 
\sim \cases{H_0^{-1} \Omega_R^{-1/2}(1+z)^{-2} & 
$(z\gg z_{\rm eq})$,\cr 
H_0^{-1} \Omega_{\rm NR}^{-1/2}(1+z)^{-3/2} & 
$(z_{\rm eq}\gg z\gg z_{\rm curv}; \ \Omega_V=0)$,\cr} 
\end{eqnarray} 
Any of these equations describes the whole (thermal) history of 
our Universe. In fact, the different contributions are just simple 
monomials in $1+z$, so that in the origin of time, that is for $z$ 
large enough, the radiation term always dominates. 
As time went on, $z$ 
went down and for a certain value of $z$ (obtained when the first 
and second terms become exactly equal), there is a transition from 
a radiation dominated to a  matter dominated  epoch, 
then to a curvature dominated one, and so on; finally, the 
cc term takes over. These transitions 
occur at the values $z_{eq}$ and $z_{curv}$, obtained by 
equating the corresponding two terms in each case: 
\begin{equation} 
 1+z_{eq}\equiv\frac{\Omega_{NR}}{\Omega_R}, \qquad \qquad 1+z_{curv} 
\equiv\frac{1}{\Omega_{NR}}-1; 
\end{equation} 
\begin{itemize} 
\item radiation dom. $\longrightarrow$ matter dom. $\longrightarrow$ 
curvature dom.  $\longrightarrow$ cc: \\ this is, in a single 
line, the whole thermal history of the Universe; 
\item $\lambda (z) = \lambda_0 (1+z)^{-1}$ $\Longrightarrow$ 
$\lambda (z) > d_H(z)$, \ for  $z > z_{\rm enter}$, which is when 
general relativistic effects become important; 
\item $\lambda(t) = \lambda_0[a(t)/a_0],$ \quad $\rho(t) = 
 \rho_0 [a_0/a(t)]^3$, those are the behaviors of the wave length 
 an the density, as time elapses; 
from these behaviors we see that the mass of non-relativistic 
matter, $M(\lambda_0)$, inside a sphere of $r=\lambda_0 /2$ is 
given by 
\begin{eqnarray} 
M&=&{4\pi\over 3} \rho(t) \left[{\lambda(t)\over 2}\right]^3 = {4\pi \over 3} 
\rho_0 \left( {\lambda_0\over 2}\right)^3 \nonumber \\ 
&=& 1.45\times 10^{11}{\rm M}_\odot 
 (\Omega_{\rm NR 
} h^2) \left( {\lambda_0\over 1\, {\rm Mpc}}\right)^3, 
\end{eqnarray} so that: 
\item the co-moving scale $\lambda_0 \approx 1$ Mpc contains a 
typical {\it galaxy} mass; \item and $\lambda_0 \approx 10$ Mpc 
contains a typical  {\it cluster} mass. 
\end{itemize} 
Moreover, 
\begin{itemize} 
\item At $z \gg z_{\rm enter}$, one needs to consider 
  General Relativity in full {\it but} only linear 
perturbation theory 
  (non-linear effects are very small). 
\item For $z \ll z_{\rm enter}$ we are in the 
 non-linear epoch {\it but} only Newtonian gravity has to be 
 taken into account. 
\item The {\it role} of general relativity calculations  is thus 
to evolve initial (linear) perturbations up to $z \simeq  z_{\rm 
enter}$ where the non-linear regime takes over. 
\end{itemize} 
And the {\it techniques} commonly employed in these processes are: 
\begin{itemize} 
\item Linear growth in the general relativistic regime. \item 
Gravitational clustering in the Newtonian theory .\item Linear 
perturbations in the Newtonian limit. \item The Zeldovich 
approximation. \item The spherical approximation, leading to the 
improved spherical collapse model. \item The use of scaling laws. Etc. 
\end{itemize} 
For details see, e.g.,  the following very good 
Refs.\cite{Peebles12,Carroll1,Padm1}

\subsection{The method of zeta-function regularization} 
 
Hawking  introduced this method\cite{hawk1} as a basic 
 tool for the regularization of infinities in QFT 
in a curved spacetime.\cite{ramond1,bd1,bos1} The idea is the 
following.\cite{hawk1} One could try to tame Quantum Gravity 
 using the 
canonical approach, by defining an arrow of time and working on 
the space-like hypersurfaces perpendicular to it, with equal time 
commutation relations. Reasons against this: \begin{romanlist} 
\item there are many topologies of the 
space-time manifold that are not a product {\bf R}$\times M_3$; 
\item such non-product topologies are sometimes very interesting; 
\item what does it mean `equal time' in the presence of 
Heisenberg's uncertainty principle? \end{romanlist} 
 
One thus turns naturally towards the path-integral approach: 
 \begin{eqnarray} <g_2, 
\phi_2, {\mathcal S}_2 | g_1, \phi_1, {\mathcal S}_1> = \int {\cal 
D} [g, \phi] \ e^{iS[g, \phi]}, \end{eqnarray} where $g_j$ denotes 
the spacetime metric, $\phi_j$ are matter fields, 
 ${\mathcal S}_j$ general spacetime surfaces (${\mathcal S}_j=M_j \cup \partial M_j$), 
 $\cal D$ a measure over all 
 possible `paths' leading from the $j=1$ to the $j=2$ values of 
 the intervening magnitudes, and $S$ is the action: 
\begin{eqnarray} S=\frac{1}{16\pi G} \int (R-2\Lambda) \sqrt{-g} 
\, d^4x +\int L_m\sqrt{-g} \, d^4x, \end{eqnarray} $R$ being the 
curvature, $\Lambda$ the cosmological constant, $g$ the 
determinant of the metric, and $L_m$ the Lagrangian of the matter 
fields. Stationarity of $S$ under the boundary conditions 
\begin{eqnarray} \left. \delta g \right|_{\partial M} =0, \qquad 
\left. \vec{n}\cdot \vec{\partial} \delta g \right|_{\partial M} 
=0,\end{eqnarray} leads to Einstein's equations: 
\begin{eqnarray} R_{ab} -\frac{1}{2} g_{ab} R+\Lambda g_{ab} = 
8\pi G T_{ab}, \end{eqnarray} $T_{ab}$ being the energy-momentum 
tensor of the matter fields, namely, \begin{eqnarray} T_{ab}= 
\frac{1}{2\, \sqrt{-g}} \frac{\delta L_m}{\delta g^{ab}}. 
\end{eqnarray} The path-integral 
formalism  provides a way to deal `perturbatively' with QFT in 
curved spacetime backgrounds.\cite{bd1} First, through a rotation 
in the complex plane one defines an Euclidean action: 
\begin{eqnarray} iS \longrightarrow -\hat{S}.\end{eqnarray} One can also 
easily introduce the finite temperature formalism by the 
substitution $t_2-t_1 =i\beta$, which yields the partition 
function 
\begin{eqnarray} Z=\sum_n e^{-\beta E_n}. \end{eqnarray} If one 
now adheres to the principle that the Feynman propagator is 
obtained as the limit for $\beta \rightarrow \infty$ of the 
thermal propagator, we have shown, some time ago,\cite{ke1} that 
the usual principal-part prescription in the zeta-function 
regularization method (to be described below) need {\it not} be 
imposed any more as an additional assumption, since it beautifully 
follows from, and thus can actually be replaced, by this more 
general (and natural) principle.\cite{ke1} 
 
Next comes the stationary phase approach (also called one-loop, or 
WKB), for calculating the path integral, which consists in 
expanding around a fixed background: 
\begin{eqnarray} g=g_0 +\bar{g}, \qquad \phi =\phi_0 +\bar{\phi}, 
\end{eqnarray} what leads to the following expansion in the Euclidean 
metric: 
\begin{eqnarray} \hat{S}[g,\phi] = 
\hat{S}[g_0,\phi_0]+S_2[\bar{g},\bar{\phi}] + \cdots 
\end{eqnarray} This is most suitably expressed in terms of determinants 
(for bosonic, resp. 
fermionic fields) of the kind (here $A,B$  are  the 
relevant (pseudo-)differential operators in the corresponding 
Lagrangian): 
\begin{eqnarray} \Delta_\phi = \det \left( \frac{1}{2\pi \mu^2} 
A\right)^{-1}, \quad \Delta_\psi = \det \left( \frac{1}{2 \mu^2} 
B\right). \end{eqnarray} 
 
\subsection{A word on determinants} 
 
Many fundamental calculations of QFT reduce, in essence, to the 
computation of the determinant of some operator. One could even 
venture to say that, at one-loop order, any such theory reduces to 
a theory of determinants.  The operators involved are 
`differential' ones, as the normal physicist would say. In fact, 
properly speaking, they are pseudodifferential operators 
($\Psi$DO), that is, in loose terms `some analytic functions of 
differential operators' (such as $\sqrt{1+D}$ or $\log (1+D)$, but 
{\it not!} $\log D$). This is explained in detail in 
Refs.\cite{elicmp,cald,psdo} 
 
Important as the concept of determinant of a differential or 
$\Psi$DO may be for theoretical physicists (in view of what has 
just been said), it is surprising that this seems not to be a 
subject of study among function analysts or mathematicians in 
general. This statement must be qualified: I am specifically 
referring to determinants that involve in its definition some kind 
of regularization, very much related to operators that are not 
trace-class. This piece of calculus ---always involving 
regularization--- falls outside the scope of the standard 
disciplines and even many physically oriented mathematicians know 
little or nothing about it. In a sense, the subject has many 
things in common with that of {\it divergent series} but has not 
been so deeply investigated and lacks any reference comparable to 
the very beautiful book of Hardy,\cite{hardy} already mentioned. 
Actually, from this general viewpoint, the question of 
regularizing infinite determinants was already addressed by 
Weierstrass in a way that, although it has been pursued by some 
theoretical physicists with success, is not without problems ---as 
a general method--- since it ordinarily leads to non-local 
contributions that cannot be given a physical meaning in QFT. We 
should mention, for completion, that there are, since long ago, 
well established theories of determinants for degenerate 
operators, for trace-class operators in the Hilbert space, 
 Fredholm operators, etc.\cite{kato} 
but, again, these definitions of determinant do not fulfill all 
the needs mentioned above which arise in QFT. 
 
Any high school student knows what a determinant is, in simple 
words, or at least how to calculate the determinant of a $3\times 
3$ matrix (and some of them, even that of a $4\times 4$ one). But 
many one prominent mathematician will answer the question: {\it 
What is your favorite definition of determinant of a differential 
operator?} with: {\it I don't have any}, or: {\it These operators 
don't have determinants!} An even more `simple' question I dare to 
ask the reader (which she/he may choose to ask to some other 
colleague on its turn) is the following: {\it What is the value of 
the determinant of minus the identity operator in an infinite 
dimensional space?} Followed by: {\it And that of the determinant 
$\prod_{n \in N} (-1)^n$?} {\it Is it actually equal to the 
product of the separate determinants of the plus 1s and of the 
minus 1s?} 
 
In this short note I will point out to specific situations, some 
of them having become common lore already and other that have 
appeared recently in the literature, concerning the concept of 
determinant in QFT, and I will try to give `reasonable' answers to 
questions such as the last ones. As already mentioned, the 
mathematical theory of divergent series has been very fruitful in 
taming the infinities that have appeared in QFT, from the very 
beginning of its conception. Its role is very essential, at least 
in the first stage of the regularization/renormalization 
procedure. Euler and Borel summation methods, and analytic 
continuation techniques are there commonly used. But some 
difficulties exist that are inherent to the theory of divergent 
series (see, for instance,\cite{hardy}). One of them is the well 
known fact that, sometimes, by using different schemes, different 
results are obtained. In a well posed physical situation, the 
`right' one can then only be chosen after experimental validation. 
Another problem is to understand, in physical terms, what you are 
doing, while performing say an analytic continuation from one 
region of the complex plane to another. This has prevented e.g. 
the zeta function regularization procedure from getting general 
acceptance among common physicists.\cite{zb1,zb2,zb3} 
 
The situation  concerning infinite determinants is even worse, in 
a sense. There is no book on the subject to be compared, for 
instance, with the above mentioned one by Hardy and we see every 
day that dubious manipulations are being performed at the level of 
the eigenvalues, that are then translated to the determinant 
itself and elevated sometimes to the category of standard results 
---when not of lore theorems. The first problem is the definition 
of the determinant itself. Let me quote in this respect from a 
famous paper by E. Witten:\cite{Witten1} {\it The determinant of 
the Dirac operator is defined roughly as \begin{equation} \det 
{\cal D} = \prod_i \lambda_i, \end{equation}where the infinite 
product is regularized with (for example) zeta function or 
Pauli-Villars regularization}. The zeta function definition of the 
determinant\begin{equation} {\det}_\zeta {\cal D} =\exp 
\left[-{\zeta_{\cal D}}' (0)\right], \end{equation} is maybe the 
one that has more firm mathematical grounds.\cite{RS} In spite of 
starting from the identity: log det $=$ tr log,  it is known to 
develop the so called {\it multiplicative anomaly}: the 
determinant of the product of two operators is not equal, in 
general, to the product of the determinants (even if the operators 
commute!). This happens already with very simple operators (as two 
one-dimensional harmonic oscillators only differing in a constant 
term, Laplacians plus different mass terms, etc.). It may look 
incredible, at first sight, from the tr log property and the 
additivity of the trace, but we must just take into account that 
the zeta trace is {\it no} ordinary trace (for it involves 
regularization), namely: 
\begin{equation} \mbox{Tr}_\zeta {\cal D} =\zeta_{\cal D} (-1), \end{equation} so 
that $\mbox{Tr}_\zeta (A+B) \neq \mbox{Tr}_\zeta A + \mbox{Tr 
}_\zeta B$, in general. Not to understand this has originated a 
considerable amount of errors in the specialized literature 
---falsely attributed to missfunctions of the rigorous and elegant 
zeta function method! 
 
As an example, consider the following commuting linear operators 
in an infinite-dimensional space, given in diagonal form 
by:\begin{equation} O_1 = \mbox{diag\ } (1,2,3,4, \ldots ), \qquad 
O_2 = \mbox{diag\ } (1,1,1,1, \ldots ) \equiv  I, \end{equation} 
and their sum \begin{equation} O_1 +O_2 = \mbox{diag\ } (2,3,4,5, 
\ldots ). \end{equation} The corresponding $\zeta$-traces are 
easily obtained: 
\begin{eqnarray} &&\mbox{Tr}_\zeta O_1 = \zeta_R (-1) = - \frac{1}{12}, 
\qquad \mbox{Tr}_\zeta O_2 = \zeta_R (0) = - \frac{1}{2}, 
\nonumber 
\\ && \mbox{Tr}_\zeta (O_1 +O_2)= \zeta_R (-1) -1= - \frac{13}{12}, 
\end{eqnarray} the last trace having been calculated according to 
the rules of infinite series summation (see e.g., Hardy\cite{hardy}). 
We observe that \begin{equation} \mbox{Tr}_\zeta 
(O_1 +O_2) - \mbox{Tr}_\zeta O_1 - \mbox{Tr}_\zeta O_2=  - 
\frac{1}{2} \neq 0. 
\end{equation} If this happens in such  simple situation, 
involving the identity operator, one can easily imagine that any 
precaution one can take in manipulating infinite sums might turn 
out to be insufficient. Moreover, since the multiplicative anomaly 
---as has been pointed out before--- originates precisely in the 
failure of this addition property for the regularized trace, we 
can already guess that it also can show up in very simple 
situations,  in fact. The appearance of the 
multiplicative anomaly prevents, in particular, naive manipulations 
with the eigenvalues in the determinant, as re-orderings and 
splittings, what a number of physicists seem not to be aware of. 
All the above considerations may sound rather trivial, but 
actually they are not,  and should be carefully taken into account 
before proceeding with the sort of manipulations of the 
eigenvalues and splittings of determinants that pervade the 
specialized literature. 
 
\subsection{The zero point energy} 
 
If $H$ is now the Hamiltonian corresponding to a physical, quantum 
system, the zero point energy is given by \begin{eqnarray} <0| 
H|0>, \end{eqnarray} where $|0>$ is the vacuum state. In general, 
after normal ordering we'll have: \begin{eqnarray} H= \left( n+ 
\frac{1}{2} \right) \, \lambda_n \, a_n\, a_n^\dagger, 
\end{eqnarray} and this yields for the vacuum energy: \begin{eqnarray} <0| 
H|0>= \frac{\hbar c}{2} \sum_n \lambda_n. \end{eqnarray} (I won't 
normally keep track of the $\hbar$'s and $c$'s that will be set 
equal to 1.) The physical meaning of this energy was the object of 
a very long controversy, involving many first-rate physicists, 
until the late Heindrik Casimir gave the explanation (over fifty 
years ago) that is widely accepted nowadays, and that's the reason 
why the zero-point energy is usually associated to his 
name.\cite{Casimir} 
 
Only in  special cases will this sum be convergent. Generically 
one gets a divergent series, to be regularized by different means. 
The zeta-function method\cite{zb1} 
---which is best suited for doing these calculations\cite{zbas1}--- 
will interprete it as the value of the zeta 
function of $H$: 
$\zeta_H (s) =  \sum_n \lambda_n^{-s}$, 
at $s=-1$ (we set $\hbar = c =1$). Generically  $\zeta_H (s)$ is 
only defined as an absolutely convergent series for Re $s > a_0$ 
($a_0$  an abscissa of convergence),\footnote{Which in general it 
won't be at $s=1$, as for the case of the Riemann zeta function.} 
but it can be continued to the whole complex plane, with the 
possible appearance of poles as only singularities. If 
  $\zeta_H (s)$ has no pole at $s=-1$ then we are done; 
if it hits a pole,  further elaboration is necessary. That the 
mathematical result one thus gets coincides with 
great precision with the {\it experimental} 
one, constitutes another clear example of {\it unreasonable 
effectiveness of mathematics}.\cite{ew1} 
 
In fact things do not turn out to be so simple. One cannot assign 
a meaning to the {\it absolute} value of the zero-point energy, 
and any physical effect is an energy difference between two 
situations, such as a quantum  field in curved space as compared 
with the same field in flat space, or one satisfying BCs on some 
surface as compared with the same in its absence, etc. This 
difference is the {\it Casimir energy}:\cite{Casimir} 
\begin{eqnarray} 
E_C = E_0^{BC} - E_0 =  \frac{1}{2} \left( \mbox{Tr } H^{BC} - 
\mbox{Tr } H \right). 
\end{eqnarray} 
And  here  a very important problem appears, which has been the 
object of intermittent but continued discussion for some time: 
imposing mathematical boundary conditions (BCs) on physical 
quantum fields turns out to be a highly non-trivial act. This was 
investigated in much detail already in a paper by Deutsch and 
Candelas a quarter of a century ago,\cite{dc79} These authors 
quantized electromagnetic and scalar fields in the region 
near an arbitrary 
smooth boundary, and calculated the renormalized vacuum 
expectation value of the stress-energy tensor, to find that the 
energy density diverges as the boundary is approached. Therefore, 
regularization and renormalization did not seem to cure the 
problem with infinities in this case and an infinite {\it 
physical} energy was obtained if the mathematical BCs were to be 
fulfilled. However, the authors argued that surfaces have non-zero 
depth, and its value could be taken as a handy (dimensional) 
cutoff in order to regularize the infinities. This approach will 
be recovered later in this paper. Just two years after Deutsch and 
Candelas' work, Kurt Symanzik carried out a rigorous analysis of 
QFT  in the presence of boundaries.\cite{ks81} Prescribing the 
value of the quantum field on a boundary means using the 
Schr\"odinger representation, and Symanzik was able to show 
rigorously that such representation exists to all orders in the 
perturbative expansion. He showed also that the field operator 
being diagonalized in a smooth hypersurface differs from the usual 
renormalized one by a factor that diverges logarithmically when 
the distance to the hypersurface goes to zero. This requires a 
precise limiting procedure and  point splitting to be applied. In 
any case, the issue was proven to be perfectly meaningful within 
the domains of renormalized QFT. In this case the BCs and the 
hypersurfaces themselves were treated at a pure mathematical level 
(zero depth) by using delta functions. 
 
Recently, a new approach to the problem has been 
postulated.\cite{bj1} BCs on a field, $\phi$, are enforced on a 
surface, $S$, by introducing a scalar potential, $\sigma$, of 
Gaussian shape living on and near the surface. When the Gaussian 
becomes a delta function, the BCs (Dirichlet here) are enforced: 
the delta-shaped potential kills {\it all} the modes of $\phi$ at 
the surface. For the rest, the quantum system undergoes a 
full-fledged QFT renormalization, as in the case of Symanzik's 
approach. The results obtained confirm those of Ref.\cite{dc79} in the 
several models studied albeit they do not seem to agree with those 
of Ref.\cite{ks81}. They are also in clear contradiction with the ones 
quoted in the usual textbooks and review articles dealing with the 
Casimir effect,\cite{cb1} where no infinite energy density when 
approaching the Casimir plates has been reported.

Too often has it been argued that  sophisticated  regularization 
methods, as the zeta-function procedure, get rid of infinities in 
 an obscure way (e.g. through analytic continuation), so that, 
contrary to what happens with  cut-offs, one cannot keep trace of 
the infinities, which are cleared up without control, leading 
sometimes to erroneous results. One cannot refute a statement of 
this kind rigorously, but it should be noted that more once (if 
not always) the discrepancies between the result obtained by using 
the zeta procedure and other ---say cut-off like--- approaches 
have been proven to emerge from a {\it misuse} of zeta 
regularization, and {\it not} to stem from the method itself. When 
employed properly, the correct results have been recovered (for a 
good number of examples, see Refs.\cite{zb1,zbas1,bp1,bpg2,zb3}).

The expression above acquires a very important meaning as soon as 
one compares different settings, e.g., one where some sort of 
boundary conditions are imposed to the vacuum (e.g., a pair of 
parallel plates, infinitely conducting, in the vacuum 
corresponding to the electromagnetic field) with another situation 
where the boundary conditions (the plates) are absent (they have 
been sent to infinity). The difference yields a physically 
observable energy. 
 
In general the sums appearing here are all divergent. They give 
rise to the most primitive, but physically meaningful, examples of 
zeta function regularization one can think of. In fact, according 
to the definitions above: 
\begin{eqnarray} <0| H|0> =\frac{1}{2} \zeta_H(-1). 
\end{eqnarray} 
 
It is important to notice that the zero-point energy is something 
one always has to keep in mind when considering any sort of 
quantum effect. Its contribution can be in some cases negligible, 
even by several orders of magnitude (as seems to be the case with 
sonoluminiscence effects), but it can be of a few percent (as in 
some laser cavity effects), or even of some $10-30 \%$ as in the 
case of several wetting phenomena of alkali surfaces by Helium. 
Not to speak of the specifically devised experiments, where it may 
account for the full result. 
 
In the case of the calculation of the value of the cosmological 
constant, it is immediate to see from the 
expressions considered before that: \begin{eqnarray} <0| 
T_{\mu\nu}|0> =\frac{\Lambda}{8\pi G} +\frac{1}{2 \, V} \sum_n 
\lambda_n, 
\end{eqnarray} where $V$ is the volume of the space manifold and 
the second term as a whole is the vacuum energy density 
corresponding to the quantum field (or fields) we are considering. 
Unless the first term (the cosmological constant), the vacuum 
energy density is not a constant (it goes as $a^{-4}$, $a$ being a 
typical cosmological length). However, this does not prevent the 
mixing of the two contributions when one considers, e.g., `the 
presently observed value of the cosmological constant'. What 
we have calculated is the second contribution for a scalar 
field of very low mass. 
 
\subsection{Quantum fluctuations of the cosmological vacuum energy} 
 
The issue of the cosmological constant has got renewed thrust from 
the observational evidence of an acceleration in the expansion of 
our Universe, initially reported by two different 
groups.\cite{perl,ries,accel1} There was some controversy on 
the reliability of the results obtained from those observations 
and on its precise interpretation, by a number of different 
reasons. Anyway, after new data has been gathered, there is presently 
reasonable consensus among the community of cosmologists that 
 there is, in fact, an acceleration, 
and that it has the order of magnitude obtained in the above 
mentioned observations.\cite{ries2,Carroll1,Carroll2} 
 In support of this consensus, the recently 
issued analysis of the data taken by the BOOMERAanG\cite{boom} and 
MAXIMA-1\cite{max1} balloons have been correspondingly crossed 
with those from the just mentioned observations, to conclude that 
the results of BOOMERanG and MAXIMA-1 can perfectly account for an 
accelerating universe and that, taking together both kinds of 
observations, one infers that we most probably live in a flat 
universe. As a consequence, many theoreticians have urged to try 
to explain this fact, and also to try to reproduce the precise 
value of the cosmological constant coming from these observations, 
in the available models.\cite{sts1,shs1,mon1} 
 
Now, as crudely stated by Weinberg in a review paper,\cite{wei2} 
it is even more difficult to explain why the cosmological constant 
is so small but non-zero, than to build theoretical models where 
it exactly vanishes.\cite{cwp1} Rigorous calculations performed in 
quantum field theory on the vacuum energy density, $\rho_V$, 
corresponding to quantum fluctuations of the fields we observe in 
nature, lead to values that are over 120 orders of magnitude in 
excess of the values allowed by observations of the space-time 
around us. 
 
Rather than trying to understand the fine-tuned cancellation of 
such enormous values at this {\it local} level (a very difficult 
question that we are going to leave unanswered, and even 
unattended, here), in this section we will elaborate on a quite 
simple and primitive idea (but, for the same reason, of far 
reaching, inescapable consequences), related with the {\it global} 
topology of the universe\cite{ct1} and in connection with the 
possibility that a very faint, massless scalar field pervading the 
universe could exist. Fields of this kind are ubiquitous in 
inflationary models, quintessence theories, and the like. In other 
words, we do not pretend  to solve the old problem of the 
cosmological constant, not even to contribute significantly to its 
understanding, but just to present an extraordinarily simple model 
which shows that the right order of magnitude of (some 
contributions to) $\rho_V$, in the precise range deduced from the 
astrophysical observations,\cite{perl,ries} e.g. $\rho_V \sim 
10^{-10}$ erg/cm$^3$, are not difficult to get. To say it in 
different words, we only address here what has been termed by 
Weinberg\cite{wei2} the {\it new} cosmological constant problem. 
 
In short, we shall assume the existence of a scalar field 
background extending through the universe and shall calculate the 
contribution to the cosmological constant coming from the Casimir 
energy density\cite{Casimir} corresponding to this field for some 
typical boundary conditions. The ultraviolet contributions will be 
safely set to zero by some mechanism of a fundamental theory. 
 Another hypothesis will be the 
existence of both large and small dimensions (the total number of 
large spatial coordinates will be always three), some of which 
(from each class) may be compactified, so that the global topology 
of the universe will play an important role, too. There is by now 
a quite extensive literature both in the subject of what is the 
global topology of spatial sections of the universe\cite{ct1} and 
also on the issue of the possible contribution of the Casimir 
effect as a source of some sort of cosmic energy, as in the case 
of the creation of a neutron star.\cite{sokol1} There are 
arguments that favor different topologies, as a compact hyperbolic 
manifold for the spatial section, what would have clear 
observational consequences.\cite{css-mfo} Other interesting work 
along these lines was reported in Ref.\cite{eejmp12} and related ideas 
have been discussed very recently in Ref.\cite{banks1}. However, we 
 differ from all those in several respects. To begin, the 
emphasis is put now in obtaining the right order of magnitude for 
the effect, e.g., one that matches the recent observational 
results. At the present stage, in view of the observational 
precision, it has no sense to consider the whole amount of 
possibilities concerning the nature of the field, the different 
models for the topology of the universe, and the different 
boundary conditions possible, with its effect on 
 the sign of the force. 
 
At this level, from our previous experience in these calculations 
and from the many tables (see, e.g., Refs.\cite{zb1,zb2,zbas1} where 
precise values of the Casimir effect corresponding to a number of 
different configurations have been reported), we realize  that the 
range of {\it orders of magnitude} of the vacuum energy density 
for the most common possibilities is not so widespread, and may 
only differ by at most a couple of digits. This will allow us, 
both for the sake of simplicity {\it and} universality, to deal 
with a most simple situation, which is the one corresponding to a 
scalar field with periodic boundary conditions. Actually, as 
explained in Ref.\cite{eenc} in detail, all other cases for parallel 
plates, with any of the usual boundary conditions, can be reduced 
to this one, from a mathematical viewpoint. 

\subsection{Two basic space-time models} 
 
Let us thus consider a universe with a space-time of one of the 
following types: $\mathbf{R^{d+1}} \times \mathbf{T}^p\times 
\mathbf{T}^q$, $\mathbf{R^{d+1}} \times \mathbf{T}^p\times \mathbf{S}^q, 
\ldots$, which are actually plausible models for the space-time 
topology. A (nowadays) free scalar field pervading the universe 
will satisfy \begin{eqnarray} (-\Box +M^2) \phi =0, \end{eqnarray} 
restricted by the 
appropriate boundary conditions (e.g., periodic, in the first case 
considered).  Here, $d\geq 0$ stands for a possible number of 
non-compactified dimensions. 
 
Recall now that the physical contribution to the vacuum or 
zero-point energy $< 0 | H |  0 >$ (where $H$ is the Hamiltonian 
corresponding to our massive scalar field and $|  0 >$ the vacuum 
state) is obtained on subtracting to these expression ---with the 
vacuum corresponding to our compactified spatial section with the 
assumed boundary conditions--- the vacuum energy corresponding to 
the same situation with the only change that the compactification 
is absent (in practice this is done by conveniently sending the 
compactification radii to infinity). As well known, both of these 
vacuum energies are in fact infinite, but it is its {\it 
difference} 
\begin{eqnarray} 
E_C = \left.  < 0 | H |  0 >\right|_R  - \left. < 0 | H | 
0 >\right|_{R\rightarrow \infty} 
\end{eqnarray} 
(where $R$ stands here  for a typical compactification length) 
that makes physical sense, giving rise to the finite value of the 
Casimir energy $E_C$, which will depend on $R$ (after a well 
defined regularization/renormalization procedure is carried out). 
In fact we will discuss the Casimir (or vacuum) energy {\it 
density}, $\rho_C=E_C/V$, which can account for either a finite or 
an infinite volume of the spatial section of the universe (from 
now on we shall assume that all diagonalizations already 
correspond to energy densities, and the volume factors will be 
replaced at the end). In terms of the spectrum $\{\lambda_n\}$ of 
$H$: 
\begin{eqnarray} 
< 0 | H |  0 > = \frac{1}{2} \sum_n \lambda_n, 
\end{eqnarray} 
where the sum over $n$ is a sum over the whole spectrum, which 
involves, in general, several continuum and several discrete 
indices. The last appear tipically when compactifying the space 
coordinates (much in the same way as time compactification gives 
rise to finite-temperature field theory), as in the cases we are 
going to consider. Thus, the cases treated will involve 
integration over $d$ continuous dimensions and multiple summations 
over $p+q$ indices (for a pedagogical description of this 
procedure, see Ref.\cite{eenc}). 
 
To be precise, the physical vacuum energy density corresponding to our case, 
where the contribution of a scalar field, $\phi$ in a (partly) compactified 
spatial section of the universe is considered, will be 
denoted by $\rho_\phi$ (note that this is just the contribution to 
$\rho_V$ coming from this field, there might be other, in general). 
It is given by \begin{eqnarray} 
 \rho_\phi =\frac{1}{2} 
\sum_{\mbox{\bf k}} \frac{1}{\mu} \left(k^2 +M^2\right)^{1/2}, 
\label{c2} \end{eqnarray} where the sum $\sum_{\mbox{\bf k}}$ is a 
generalized one (as explained above) and $\mu$ is the usual 
mass-dimensional parameter to render the eigenvalues adimensional 
(we take $\hbar =c =1$ and shall insert the dimensionfull units 
only at the end of the calculation). The mass $M$ of the field 
will be here considered to be arbitrarily small and will be kept 
different from zero, for the moment, for computational reasons 
---as well as for physical ones, since a very tiny mass for the 
field can never be excluded. Some comments about the choice of our 
model are in order. The first seems obvious: the coupling of the 
scalar field to gravity should be considered. This has been done 
in all detail in, e.g., Ref.\cite{pr1} (see also the references 
therein). The conclusion is that taking it into account does not 
change the results to be obtained here. Of course, the 
renormalization of the model is rendered much more involved, and 
one must enter a discussion on the orders of magnitude of the 
different contributions, which yields, in the end, an ordinary 
perturbative expansion, the coupling constant being finally 
re-absorbed into the mass of the scalar field. In conclusion, we 
would not gain anything new by taking into account the coupling of 
the scalar field to gravity. Owing, essentially, to the smallness 
of the resulting mass for the scalar field, one can prove that, 
quantitatively, the difference in the final result is at most of a 
few percent. 
 
Another important consideration is the fact that our model is 
stationary, while the universe is expanding. Again, careful 
calculations show that this effect can actually be dismissed at 
the level of our order of magnitude calculation, since its value 
cannot surpass the one that we will get (as is seen from the 
present value of the expansion rate $\Delta R /R \sim 10^{-10}$ 
per year or from direct consideration of the Hubble coefficient). 
 As  before, for the sake of simplicity, and in order to focus just on 
 the essential issues of our argument, we will perform 
a (momentaneously) static calculation. As a consequence, the value 
of the Casimir energy density, and of the cosmological constant, 
to be obtained will correspond to the present epoch, and are bound 
to change with time. 
 
The last comment at this point would be that (as shown by the many 
references mentioned above), the idea presented here is not 
entirely new. However, the simplicity and the generality of its 
implementation below are indeed new. The issue at work here 
is absolutely independent of {\it any} specific model, the only 
assumptions having been clearly specified before (e.g., existence 
of a very light scalar field and of some reasonably compactified 
scales, see later). Secondly, it will turn out, in the end, that 
the only `free parameter' to play with (the number of compactified 
dimensions) will actually not be that `free' but, on the contray, 
very much constrained to have an admissible value. This will 
become clear after the calculations below. Thirdly, although the 
calculation may seem easy to do, in fact it is not so. Some 
 reflection identities, due to the author, will allow 
 to be performed  analitically.

\subsection{The vacuum energy density and its regularization}

To exhibit explicitly a couple of the wide family of cases 
considered, let us write down in detail  the formulas 
corresponding to the two first topologies, as described above. For 
a ($p,q$)-toroidal universe, with $p$ the number of `large' and 
$q$ of `small' dimensions: \begin{eqnarray} \rho_\phi 
&=&\frac{\pi^{-d/2}}{2^d\Gamma (d/2) \prod_{j=1}^p a_j 
\prod_{h=1}^q b_h} \int_0^\infty dk \, k^{d-1} \sum_{\mbox{\bf 
n}_p=-\mathbf{\infty}}^{\mathbf{\infty}} \sum_{\mbox{\bf 
m}_q=-\mathbf{\infty}}^{\mathbf{\infty}} \nonumber \\ 
&& \left[ \sum_{j=1}^p \left( \frac{2\pi n_j}{a_j}\right)^2 + 
\sum_{h=1}^q\left( 
\frac{2\pi m_h}{b_h}\right)^2 +M^2 \right]^{1/2} \label{t1} \nonumber \\ 
&& \sim  \frac{1}{a^pb^q} \sum_{\mbox{\bf n}_p, \mbox{\bf 
m}_q=-\mathbf{ \infty}}^{\mathbf{\infty}} \left( 
\frac{1}{a^2}\sum_{j=1}^p n_j^2 + \frac{1}{b^2} \sum_{h=1}^q m_h^2 
+M^2 \right)^{(d+1)/2+1},\label{t2} 
\end{eqnarray} where the last formula corresponds to the case when 
all large (resp. all small) compactification scales are the same. 
In this last expression the squared mass of the field should be 
divided by $4\pi^2\mu^2$, but we have renamed it again $M^2$ to 
simplify the ensuing formulas (as $M$ is going to be very small, 
we need not keep track of this change). We also will not take care 
for the moment of the mass-dim factor $\mu$ in other places $-$as 
is usually done$-$ since formulas would get unnecesarily 
complicated and there is no problem in recovering it at the end of 
the calculation. For a ($p$-toroidal, $q$-spherical)-universe, the 
expression turns out to be \begin{eqnarray} \rho_\phi 
&=&\frac{\pi^{-d/2}}{2^d\Gamma (d/2)\, \prod_{j=1}^p a_j \ b^q} 
\int_0^\infty dk \, k^{d-1}\, \sum_{\mbox{\bf 
n}_p=-\mathbf{\infty}}^{\mathbf{\infty}} \sum_{l=1}^\infty 
P_{q-1}(l)\nonumber \\ 
&& \times \left[ \sum_{j=1}^p \left( \frac{2\pi n_j}{a_j}\right)^2 
+ \frac{Q_2(l)}{b^2} +M^2 \right]^{1/2} 
    \label{ts1} \nonumber \\ &\sim & \frac{1}{a^pb^q} \sum_{\mbox{\bf 
n}_p=-\mathbf{\infty}}^{\mathbf{\infty}} \sum_{l=1}^\infty 
P_{q-1}(l)  \left( \frac{4\pi^2}{a^2}\sum_{j=1}^p n_j^2 + 
 \frac{l(l+q)}{b^2}  +M^2 \right)^{(d+1)/2+1},\label{ts2} 
\end{eqnarray} where $P_{q-1}(l)$ is a polynomial in $l$ of degree $q-1$, 
and where the second formula corresponds to the similar situation 
as the second one before. On dealing with our observable universe, 
in all these expression we assume that $d=3-p$, the number of 
non-compactified, `large' spatial dimensions (thus, no $d$ 
dependence will remain). 
 
As is clear, all these expressions for $\rho_\phi$ need to be 
regularized. We will use zeta function regularization, taking 
advantage of the very powerful equalities that have been derived 
by the author,\cite{eecmp1,ke1} and 
which reduce the enormous burden of 
such computations to the easy application of some formulas. For 
the sake of completeness, let us very briefly summarize how this 
works.\cite{eejpa1,eenc} We deal here only with the case when the 
spectrum of the Hamiltonian operator is known explicitly. Going 
back to the most general expressions of the Casimir energy 
corresponding to this case, namely Eq. (\ref{c2}), we replace the 
exponents in them with a complex variable, $s$, thus obtaining the 
zeta function associated with the operator as: 
\begin{eqnarray} \zeta (s) =\frac{1}{2} \sum_{\mbox{\bf k}} 
\left(\frac{k^2 +M^2}{\mu^2}\right)^{-s/2}. \label{z2} 
\end{eqnarray} The next step is to perform the analytic 
continuation of the zeta function from a domain of the complex 
$s$-plane with Re $s$ big enough (where it is perfectly defined by 
this sum) to the point $s=-1$, to obtain: 
\begin{eqnarray} \rho_\phi = \zeta (-1). \end{eqnarray} The effectiveness of this method 
has been sufficiently described before (see, e.g., 
\cite{zb1,zb2}). As we know from precise Casimir calculations in 
those references, no further subtraction or renormalization is 
needed in the cases here considered, in order to obtain the 
physical value for the vacuum energy density (there is actually a 
subtraction at infinity taken into account, as carefully described above, 
   but it is of null value, and 
no renormalization, not even a finite one, very common to other 
frameworks, applies here). 
 
Using the formulas\cite{eecmp1} that generalize the 
well-known Chowla-Selberg expression to the situations considered 
above, Eqs. (\ref{t1}) and (\ref{ts1}) ---namely, 
multidimensional, massive cases--- we can provide arbitrarily 
accurate results for different values of the compactification 
radii. However, as argued above we only aim here at matching 
the {\it order of magnitude} of the Casimir value and, thus, we 
shall just deal with the most simple cases of Eqs. (\ref{t2}) or 
(\ref{ts2}), 
which yield the same orders of magnitude as the rest of them). 
Also in accordance with this observation, we notice that among the 
models here considered and which lead to the values that will be 
obtained below, there are in particular the very important typical 
cases of isotropic universes with the spherical topology. As all 
our discussion here is in terms of orders of magnitude and not of 
precise values with small errors, all these cases are included on 
{\it equal footing}. But, on the other hand, it has no sense to 
present a lengthy calculation dealing in detail with all the 
possible spatial geometries. Anyhow, all these calculations 
can indeed be done, and are 
very similar to the one here, as has been 
described in detail elsewhere.\cite{eejmp12,zb1,zb2} 
 
For the analytic continuation of the zeta function corresponding 
to (\ref{t1}), we obtain:\cite{eecmp1} 
\begin{eqnarray} 
\zeta(s)&=&\frac{2\pi^{s/2+1}}{a^{p-(s+1)/2}b^{q-(s-1)/2} \Gamma 
(s/2)} \sum_{\mbox{\bf m}_q=-\mathbf{\infty}}^{\mathbf{\infty}} 
\sum_{h=0}^p \hspace*{-1mm}\left(_{\,\displaystyle 
h\,}^{\,\displaystyle p\,}\right) 2^h 
\hspace*{-2mm}\sum_{\mbox{\bf n}_h=1}^{\mathbf{\infty}} 
\hspace*{-2mm}\left( \frac{\sum_{j=1}^h n_j^2 }{\sum_{k=1}^q 
m_k^2+M^2}\right)^{(s-1)/4} \nonumber \\ && \times K_{(s-1)/2} 
\left[ \frac{2\pi a}{b} \sqrt{\sum_{j=1}^h n_j^2 
\left(\sum_{k=1}^q m_k^2+M^2\right)}\right], \label{z11} 
\end{eqnarray} where $K_\nu (z)$ is the modified Bessel function 
of the second kind. Having performed already the analytic 
continuation, this expression is ready for the substitution 
$s=-1$, and yields 
 \begin{eqnarray} 
\rho_\phi &=& -\frac{1}{a^pb^{q+1}} \sum_{h=0}^p 
\left(_{\,\displaystyle h\,}^{\,\displaystyle p\,}\right) 2^h 
\sum_{\mbox{\bf n}_h=1}^{\mathbf{\infty}} \sum_{\mbox{\bf 
m}_q=-\mathbf{\infty}}^{\mathbf{\infty}} \sqrt{\frac{\sum_{k=1}^q 
m_k^2+M^2}{\sum_{j=1}^h n_j^2 }} \nonumber \\ 
&& \times K_1 \left[ \frac{2\pi a}{b} \sqrt{\sum_{j=1}^h n_j^2 
\left(\sum_{k=1}^q m_k^2+M^2\right)}\right]. \label{c11} 
\end{eqnarray} Now, from the behaviour of the function $K_\nu (z)$ 
for small values of its argument, \begin{eqnarray} K_\nu (z) \sim 
\frac{1}{2} \Gamma (\nu) (z/2)^{-\nu}, \qquad z \to 0, \label{kna} 
\end{eqnarray} we obtain, in the case when $M$ is very small, 
\begin{eqnarray} \rho_\phi &=& -\frac{1}{a^pb^{q+1}} \left\{ M\, K_1 \left( 
\frac{2\pi a}{b} M \right)+ \sum_{h=0}^p \left(_{ \, \displaystyle 
h\,}^{\, \displaystyle p\,}\right) 2^h \sum_{\mbox{\bf 
n}_h=1}^{\mathbf{\infty}} \frac{M}{\sqrt{\sum_{j=1}^h n_j^2 }} 
\right.\nonumber \\ && \times K_1\left( \frac{2\pi a}{b} M 
\sqrt{\sum_{j=1}^h n_j^2} \right) + \left. {\cal O} \left[ 
q\sqrt{1+M^2} K_1\left( \frac{2\pi a}{b}\sqrt{1+M^2}\right) 
\right]\right\}. \label{ff0}\end{eqnarray} At this stage, the only 
presence of the mass-dim parameter $\mu$ is as $M/\mu$ everywhere. 
This does not conceptually affect the small-$M$ limit, $M/\mu << 
b/a$. Using (\ref{kna}) and inserting now in the expression the 
$\hbar$ and $c$ factors, we finally get \begin{eqnarray} \rho_\phi 
= -\frac{\hbar c}{2\pi a^{p+1}b^q} \left[1+\sum_{h=0}^p 
\left(_{\,\displaystyle h\,}^{\,\displaystyle p\,}\right) 2^h 
\alpha \right]+ {\cal O} \left[ q K_1\left( \frac{2\pi 
a}{b}\right) \right], \label{ff1} \end{eqnarray} where $\alpha$ is 
some finite constant, computable and under control, which is 
obtained as an explicit geometrical sum in the limit $M\rightarrow 
0$. It is remarkable that we here obtain such a well defined 
limit, independent of $M^2$, provided that $M^2$ is small enough. 
In other words, a physically very nice situation turns out to 
correspond, precisely, to the mathematically rigorous case. This 
is moreover, let me repeat, the kind of expression that one gets 
not just for the model considered, but for {\it many} other 
cases, corresponding to different fields, topologies, and boundary 
conditions ---aside from the sign in front of the formula, that 
may change with the number of compactified dimensions and the 
nature of the boundary conditions (in particular, for Dirichlet 
boundary conditions one obtains a value in the same order of 
magnitude but of opposite sign). 
 
\subsection{Numerical results}

For the most common variants, the constant $\alpha$ in (\ref{ff1}) 
has been calculated to be of order $10^2$, and the whole factor, 
in brackets, of the first term in (\ref{ff1}) has a value of order 
$10^7$. This shows the value of a precise calculation, as the one 
undertaken here, together with the fact that just a naive 
consideration of the dependencies of $\rho_\phi$ on the powers of 
the compactification radii, $a$ and $b$, is {\it not enough} in 
order to obtain the correct result. Notice, moreover, the 
non-trivial change in the power dependencies from going from Eq. 
(\ref{ff0}) to Eq. (\ref{ff1}). 
 
For the compactification radii at small scales, $b$, we shall 
simply take the magnitude of the Planck length, $b \sim 
l_{P(lanck)}$, while the typical value for the large scales, $a$, 
will be taken to be the present size of the observable universe, 
$a\sim R_U$. With this choice, the order of the quotient $a/b$ in 
the argument of $K_1$ is as big as: $a/b \sim 10^{60}$. Thus, we 
see immediately that, in fact, the final expression for the vacuum 
energy density is completely independent of the mass $M$ of the 
field, provided this is very small (eventually zero). In fact, 
since the last term in Eq. (\ref{ff1}) is exponentially vanishing, 
for large arguments of the Bessel function $K_1$, this 
contribution is zero, for all practical purposes, what is already 
a very nice result. Taken in ordinary units (and after tracing 
back all the transformations suffered by the mass term $M$) the 
actual bound on the mass of the scalar field is   $M \leq 1.2 
\times 10^{-32}$ eV, that is, physically zero, since it is lower 
by several orders of magnitude than any bound coming from the more 
usual SUSY theories $-$where in fact scalar fields with low masses 
of the order of that of the lightest neutrino do show 
up,\cite{shs1} which may have observable implications.

\begin{table}[ht] 
\tbl{Orders of magnitude of the vacuum energy density 
contribution, $\rho_\phi$, of a massless scalar field to the 
cosmological constant, $\rho_V$, for $p$ large compactified 
dimensions and $q=p+1$ small compactified dimensions, 
$p=0,\ldots,3$, for different values of the small compactification 
length, $b$, proportional to the Planck length $l_P$. In brackets 
are the values that exactly match the observational value of the 
cosmological constant, and in parenthesis the otherwise closest 
approximations to that value.}{\large \label{tcc1} 
\begin{tabular}{|c||c|c|c|c|} 
\hline \hline $\rho_\phi$ & $p=0$ & $p=1$ & $p=2$ & $p=3$ \\ 
 \hline \hline 
$b=l_P$ & $10^{-13}$ & $10^{-6}$ & 1 & $10^5$ \\ \hline $b=10\, 
l_P$  &$10^{-14}$ & ($10^{-8}$) & $10^{-3}$ & $10$ 
 \\ \hline 
$b=10^2 l_P$  &$10^{-15}$ & [$10^{-10}$] & $10^{-6}$ & $10^{-3}$ 
 \\ \hline 
$b=10^3 l_P$ & $10^{-16}$ & ($10^{-12}$) & [$10^{-9}$] & 
$10^{-7}$ \\ \hline $b=10^4 l_P$ & $10^{-17}$ & $10^{-14}$ & 
($10^{-12}$) & [$10^{-11}$] \\ \hline $b=10^5 l_P$ & $10^{-18}$ & 
$10^{-16}$ & $10^{-15}$  & $10^{-15}$ 
 \\ \hline \hline \end{tabular}} 
\end{table} 
 
By replacing all these values in Eq. (\ref{ff1}), we obtain the 
results listed in  Table \ref{tcc1}, for the orders of magnitude 
of the vacuum energy density corresponding to a sample of 
different numbers of compactified (large and small) dimensions and 
for different values of the small compactification length in terms 
of the Planck length. Notice again that the total number of large 
space dimensions is three, as corresponds to our observable 
universe. 
 As we see from Table \ref{tcc1}, good coincidence 
with the observational value for the cosmological constant is 
obtained for the contribution of a massless scalar field, 
$\rho_\phi$, for $p$ large compactified dimensions and $q=p+1$ 
small compactified dimensions, $p=0,\ldots,3$, and this for values 
of the small compactification length, $b$, of the order of 100 to 
1000 times the Planck length $l_P$ (what is actually a very 
reasonable conclusion, according also to other approaches). 
 
To be 
noticed is the fact that full agreement is obtained only for cases 
where there is exactly one small compactified dimension in excess 
of the number of large compactified dimensions. We must point out 
that the $p$ large and $q$ small dimensions are not all that are 
supposed to exist (in that case $p$ should be at least, and at 
most, 3 and the other cases would lack any physical meaning). In 
fact, as we have pointed out before, $p$ and $q$ refer to the 
compactified dimensions only, but there may be other, 
non-compactifed dimensions (exactly $3-p$ in the case of the 
`large' ones), what translates into a slight modification of the 
formulas above,  but does not change the order of magnitude of the 
final numbers obtained, assuming the most common boundary 
conditions for the non-compactified dimensions (see e.g.\cite{zb2} 
for an explanation of this technical point). In particular, the 
cases of pure spherical compactification and of mixed toroidal 
(for small magnitudes) and spherical (for big ones) 
compactification can be treated in this way and yield results in 
the same order of magnitude range. Both these cases correspond to 
(observational) isotropic spatial geometries. Also to be remarked 
again is the non-triviality of these calculations, when carried 
out exactly, as done here, to the last expression, what is 
apparent from the use of the generalized Chowla-Selberg formula. 
Simple power counting is absolutely unable to provide the correct 
order of magnitude of the results. 
 
Dimensionally speaking, within the global approach adopted in the 
present paper everything is dictated, in the end, by the two basic 
lengths in the problem, which are its Planck value and the radius 
of the observable Universe. Just by playing with these numbers in 
the context of this precise calculation of the Casimir 
effect, we have shown that the observed value of $\rho_V$ may be 
remarkably well fitted, under general hypothesis, for the most 
common models of the space-time topology. Notice also that the 
most precise fits with the observational value of the cosmological 
constant are obtained for $b$ between $b=100 \, l_P$ and $b=1000 
\, l_P$, with (1,2) and (2,3) compactified dimensions, 
respectively. The fact that the value obtained for the 
cosmological constant is so sensitive to the input may be viewed 
as a drawback but also, on the contrary, as a very {\it positive} 
feature of our model. For one, the Table \ref{tcc1} has a sharp 
discriminating power. In other words, there is in fact no tuning 
of a `free parameter' in our model and the number of large 
compactified dimensions could have been fixed beforehand, to 
respect what we know already of our observable universe.

 Also, it proves that the observational 
value is not easy at all to obtain. Table \ref{tcc1} itself proves 
that there is only very little chance of getting the right figure 
(a truly narrow window, since very easily we are off by several 
orders of magnitude). In fact, if we trust this value with the 
statistics at hand, we can undoubtedly claim $-$through use of the 
model$-$ that the ones so clearly picked up by Table \ref{tcc1} 
are {\it the} only two  possible configurations of our observable 
universe (together with a couple more coming from corresponding 
spherical compactifications). And all them correspond to a 
marginally closed universe, in full agreement too with other 
completely independent analysis of the observational 
data.\cite{Carroll1,perl,ries} 
 
Many questions may be posed to the simple models presented here, 
as concerning the dynamics of the scalar field, its couplings with 
gravity and other fields, a possible non-symmetrical behaviour 
with respect to the large and small dimensions, or the relevance 
of vacuum polarization (see Ref.\cite{sts2} concerning this last 
point). Above we have already argued that they can be proven to 
have little influence on the final numerical result (cf., in 
particular, the mass obtained for the scalar field in Ref.\cite{pr1}, 
extremely close to our own result, and the 
corresponding discussion there). From the very existence and 
specific properties of the cosmic microwave radiation (CMB) 
$-$which mimics somehow the situation described (the `mass' 
corresponding to the CMB is also in the sub-lightest-neutrino 
range)$-$ we are led to the conclusion that such a field 
could be actually present, unnoticed, in our observable universe. 
In fact, the existence of scalar fields of very low masses is also 
demanded by other frameworks, as SUSY models, where the scaling 
behaviour of the cosmological constant has been 
considered.\cite{shs1} 
 
Let us finally recall again 
that the Casimir effect is an ubiquitous phenomena. 
Its contribution may be small (as it seems to be the case, yet controverted, 
to sonoluminiscence), of some 10-30$\%$ (that is, of the right order of 
magnitude, as in wetting phenomena involving He in condensed matter physics). Here we have seen that it provides a contribution of the 
right order of magnitude, corresponding to our present epoch in the evolution 
of the universe. The implication that this calculation bears for the early 
universe and inflation is not clear from the final result, since it should be 
adapted to the situation and boundary conditions corresponding to those 
primeval epochs, what cannot be done straightforwardly. 
Work along this line is in progress.

\section*{Acknowledgments} 
 
The author is indebted to their former students and collaborators on 
some of the subjects of these lectures: Enrique Gazta{\~n}aga, 
Pablo Fosalba, Sergi R. Hildebrandt and Jose Barriga, as well as to 
the members of the Mathematics Department, MIT, and specially to 
Dan Freedman, for the continued hospitality. This investigation 
has been supported by DGICYT (Spain), project BFM2003-00620 and by 
the APM Service, MEC (Spain), grant PR2004-0126. 
 

\end{document}